\RequirePackage{fix-cm}
\documentclass[smallextended]{svjour3}
\smartqed

\usepackage{graphicx}

\usepackage{color}
\usepackage{verbatim}
\usepackage{amssymb, amsmath}
\usepackage{fancyvrb}
\usepackage[fancyvrb]{listings}
\usepackage{ulem}
\usepackage{todonotes}
\usepackage{hyperref}
\usepackage{changebar}
\usepackage{comment}

\DeclareMathOperator*{\argmax}{arg\,max}

\begin{document}

\title{Formalising the Foundations of Discrete Reinforcement Learning in Isabelle/HOL
}

\titlerunning{Formalising the Foundations of Discrete Reinforcement Learning in Isabelle/HOL}        %

\author{Mark Chevallier         \and
        Jacques Fleuriot
}

\institute{Mark Chevallier \at
           \email{s1742487@ed.ac.uk}           
           \and
           Jacques Fleuriot \at 
           \email{jdf@ed.ac.uk} \\ 
           \\
           Artificial Intelligence and its Applications Institute (AIAI), School of Informatics, University of Edinburgh, EH8 9AB, United Kingdom
}

\maketitle

\begin{abstract}
We present a formalisation of finite Markov decision processes with rewards in the Isabelle theorem prover. We focus on the foundations required for dynamic programming and the use of reinforcement learning agents over such processes. In particular, we derive the Bellman equation from first principles (in both scalar and vector form), derive a vector calculation that produces the expected value of any policy $p$, and go on to prove the existence of a universally optimal policy where there is a discounting factor less than one. Lastly, we prove that the value iteration and the policy iteration algorithms work in finite time, producing an $\epsilon$-optimal and a fully optimal policy respectively.
\end{abstract}

\section{Introduction}
\label{intro}
Mathematical models of real-life problems are frequently useful in providing insights into developing strategies to deal with those problems. Markov decision processes \cite{white1989markov} (commonly referred to as MDPs) are especially useful models for this purpose when it comes to certain classes of problems.\

MDPs are used to model environments in which an agent can act over a discrete set of states by taking particular actions in discrete time. An agent taking an action $a$ in a state $s$ can then probabilistically transition into a subsequent state $s'$ that only depends on the specific MDP and how it models the initial state and the action taken.\

Importantly, we insist that an MDP has the Markov property: the result of moving between time steps $t$ and $t+1$ depend only on the conditions and actions taken at time step $t$; the history of the agent and the MDP before that is irrelevant. Thus the MDP is memoryless.\

Taking a particular action and obtaining a result from it gives a reward to the agent. In general, the aim of modelling an environment in this way is to suppose that by optimising the reward gained, we optimise an agent's behaviour. An agent negotiating the MDP model behaves optimally if it acts in such a way as to maximise its reward over time.\

MDPs were developed from optimisation theory \cite{bellman1957markovian} and have practical applications in many fields, including medical, financial and logistical \cite{alagoz2010markov,bauerle2011markov,tetteroo2019c2c,white1985real}. In more recent years, they have frequently been used as environments over which a reinforcement learning agent or other autonomous agent can act \cite{sigaud2013markov,watkins1989learning}. Many reinforcement learning approaches that are over discrete states and times use MDPs, and having a formal model with rigorously proven properties is the first step to any formalisation of these methods.\

There is an existing formalisation of MDPs and their precursors, Markov chains (where no choice of action is permitted and instead agents simply transition from one state to another probabilistically \cite{norris1998markov}) for Isabelle by H\"olzl \cite{holzl2017markov}. In this work, though, we provide an alternative formalisation which we extend to consider the general concepts of reward needed to show the fundamentals of MDP optimisation. We will discuss this further in section \ref{sec:Background}, where we briefly examine the mathematical background to MDPs and review the existing formalisation and why we believe it isn't suitable for our goals. We also provide a quick introduction to Isabelle sufficient for our presentation in this section. The rest of the paper is then organised as follows.\

In section \ref{sec:fundform}, we introduce the specifics of our model and our early proofs based on it. Our aim here is to demonstrate that this meets the mathematical expectations of an MDP, and we do so by deriving the Bellman equation.\

In section \ref{sec:proofopt}, we discuss a pen-and-paper proof of the existence of an optimal choice of actions for an agent (a \textit{policy}) on any MDP and its mechanisation. We discuss some specific issues we had finding appropriate type representation in Isabelle and how we resolved these difficulties. Finally, we explain how we built our formal proof of the existence of an optimal policy on our model of MDPs. In doing so, we discovered some weaknesses in the pen-and-paper proof we were using as our basis and we disclose how we addressed them in the context of our finite MDPs.\

In section \ref{sec:iterations}, we conclude the current work on our model by looking at value and policy iterations. These are algorithms intended to provide a means to compute the optimal policy (or an arbitrarily close approximation) in finite time. We examine the steps for each of the algorithms and then formally prove that both work as intended.

We conclude with a summary of the work done and a discussion of possible future directions.

\section{Background}
\label{sec:Background}

We introduce Markov decision processes by their mathematical description and then go on to give a brief discussion of Isabelle and of the existing formalisation of MDPs within it.\

\subsection{Markov decision processes}

A finite Markov decision process can be considered as a tuple $$(S,A,P,R)$$ where:\

\begin{itemize}
    \item $S$ is a finite set of states representing differing conditions in the environment. States can be considered ``terminating'' if no actions are possible from them.
    \item $A(s)$ is a function on the states which returns a finite set of possible actions that can be taken from state $s$.
    \item $P(s,a,s')$ is a function which gives the probability of an agent transitioning from state $s$ into state $s'$ if it performs action $a$. Each choice of action gives a probability distribution over the states, so we have: 
    $$\forall s,a. \left(\sum_{s' \in S} P(s,a,s')\right) = 1$$ 
    $$\forall s,a,s'. P(s,a,s') \geq 0$$
    \item $R(s,s',a)$ is a function which gives the reward for transitioning from state $s$ to state $s'$ by performing action $a$. Rewards are real valued.
\end{itemize}

The goal for an agent navigating an MDP is to maximise its total reward earned. Naturally, given there is no restriction on $R(s,s',a)$, we cannot guarantee that the rewards will converge for an arbitrary MDP if there is no finite cap on the number of actions an agent may perform, which is our assumption throughout this paper and is commonly called an infinite horizon \cite{puterman2014markov}.\

Consequently, it is common to introduce a discounting factor $\gamma$, where $0\leq\gamma\leq1$ (note that if $\gamma=1$ this is the same as having no discounting factor) \cite{puterman2014markov}. Its effect is that at time step $n$, any rewards earned from time step $n+1$ onwards are multiplied by $\gamma$ before evaluation, and this discounting is compounded for future time steps. Its purpose is to enable us to estimate total future rewards by introducing a mechanism by which we can guarantee convergence for any MDP that offers bounded rewards. We can interpret it as meaning that we value distant future rewards as being worth less to us than immediate rewards.\

If we index the state of an agent at time step $t$ as $s_t$, and its actions at $t$ as $a_t$, the total discounted value of the agent performing a choice of actions from an initial state $s_0$ would therefore be: $$R(s_0,s_1,a_0) + \gamma R(s_1,s_2,a_1) + ... + \gamma^{t} R(s_t,s_{t+1},a_t) + ...$$

Normally when we consider an agent navigating an MDP, we view it as following a particular policy $\pi$ that chooses the action for the agent to perform in each state. Thus, $\pi(s)$ is a function on the states where $\forall s.\pi(s) \in A(s)$.\

When we want to evaluate potential policies to estimate the rewards an agent might earn from pursuing them, we use the concept of value, which is the expected rewards accounting for the discount. Thus, $V(s,\pi)$ is a function of states and policies that gives us the total expected discounted future reward for an agent pursuing policy $\pi$ from state $s$ -- we call this the $V$ value function.\ 

\label{Qmathdef}
Similarly, $Q(s,a,\pi)$ is the total discounted expected future reward for an agent choosing action $a$ from state $s$ and subsequently pursuing policy $\pi$ -- the $Q$ value is distinct from the $V$ value by its allowance that the initial action taken may differ from $\pi(s)$. Thus, the $V$ value is a special case of the $Q$ value.\

Obviously, these can only be evaluated when we can be sure that the sum of rewards will converge, so this condition is typically met by having $\gamma<1$. It is possible as well to ensure convergence by having a finite horizon limiting the number of steps an agent can perform, or by designing an MDP that inevitably ends in some terminating state.\

The value function is typically depicted in MDP literature using the Bellman equation \cite{barron1989bellman}, which evaluates it recursively (recall that $V$ is a special case of $Q$ where the action is chosen by the policy): 

\begin{align}
Q(s,a,\pi) = \sum_{s'} P(s,a,s') \left[ R(s,a,s') + \gamma V(s',\pi) \right]
\end{align}

To summarise the equation, recall that $Q(s,a,\pi)$ is the total expected future reward, with discounting, for taking action $a$ in state $s$ and \textit{subsequently} pursuing policy $\pi$. $P(s,a,s')$ is the probability of transitioning to state $s'$ by taking action $a$ from state $s$. $R(s,s',a)$ is the reward earned for transitioning from state $s$ to $s'$ by performing action $a$. The discount factor is given by $\gamma$ and, finally, $V(s,\pi)$ is the total expected future discounted reward for pursuing policy $\pi$ from state $s$, and is equal to $Q(s,\pi(s),\pi)$.\

\subsection{Isabelle/HOL}

We now briefly review some aspects of Isabelle/HOL, a higher order logic proof assistant, \cite{nipkow2002isabelle}, which we use to represent our model and prove its properties. Mathematical theories written in Isabelle are a collection of formal definitions of various kinds (algebraic objects, types, functions, etc.), and theorems that prove properties against them.\ 

We write our proofs in the structured proof language Isar, which depicts proofs in human-readable form \cite{wenzel1999isar} and where steps in a proof typically follow those that would be used in a naturally written mathematical proof. 

Isabelle/HOL is a typed logic that supports type classes \cite{wenzel2005using}, namely assertions of properties that can be applied against type variables to constrain and interpret them as abstract algebraic concepts: for example, as a vector or metric space. These are then known as {\it sort} constraints. If we assert that a type variable fulfils a sort constraint in a theorem, then that theorem is true for any concrete type that meets this constraint. This means, for instance, that one can assert that \texttt{int}, the type of the integers, forms a group under a given definition for the group operator by proving that it is an instance of the \texttt{group} type class. Then theorems which are true for all groups become available for the integers under this interpretation.\

When defining a constant or function, it must be assigned a type (``\texttt{t::$\tau$}'' states that \texttt{t} is of type \texttt{$\tau$}). Function types are stated using ``\texttt{$\Rightarrow$}'', from the type on the left-hand of the arrow to the type on the right-hand. Type variables are written with a prefixed apostrophe, like ``\texttt{'s}''. Sort constraints can be specified against types using suffixed ``\texttt{::}'' notation, followed by the required constraints. We provide some quick illustrative examples:

\begin{enumerate}
    \item \texttt{test\_sequence :: "nat $\Rightarrow$ real"} tells us that \texttt{test\_sequence} is a function from the \texttt{nat} type, the natural numbers including zero, to the real numbers -- a sequence of reals, in other words.
    \item \texttt{location :: "real $\Rightarrow$ real $\Rightarrow$ 'l"} tells us that \texttt{location} is a function of two real numbers which maps to a type variable \texttt{'l}.
    \item \texttt{state\_space :: "'s::\{finite\} set"} tells us that \texttt{state\_space} is a set of values of type \texttt{'s}, which we assert is finite using the Isabelle sort constraint \texttt{finite}.
\end{enumerate}

Within Isabelle, one can create a context within whose scope certain assumptions are held and certain constants have a declared type and fixed value. This is called a {\it locale} \cite{kammuller1999locales}. All theorems proven within the locale depend upon and have access to its assumptions and declared constants. As with any set of assumptions, care must be taken to ensure consistency; however, the encapsulation provided by a locale ensures that an inconsistency within its assumptions can never propagate to Isabelle's top level. An example locale definition is as follows:\

\begin{lstlisting}[basicstyle=\footnotesize\ttfamily, mathescape = true]{}
locale Sample_Locale = 
  fixes sequence :: "nat $\Rightarrow$ real"
    and $\phi$ :: real
  assumes sample_assumption: "sequence 0 = $\phi$"
\end{lstlisting}

This locale fixes two parameters: \texttt{sequence} is a function from the natural numbers to the reals and \texttt{$\phi$} is a real number. It then specifies an assumption that the value of the sequence at index 0 is $\phi$ by having \texttt{sequence 0} = $\phi$.

We rely on Isabelle's extensive library of existing work. In particular, our formalisation makes use of several existing theories from the Isabelle standard library extended by the Archive of Formal Proof (AFP) \cite{mackenzie2021evaluation}, on probability, linear algebra and analysis, which we will introduce and discuss as they arise in the rest of this paper.\

\subsection{H\"olzl's formalisation of MDPs}
\label{subs:holzl}

As previously mentioned, we are indebted to the work by H\"olzl on Markov Decision Processes \cite{holzl2017markov}, which informed the current formalisation.\ 

H\"olzl's work was performed with an emphasis on proving properties regarding the traces that an agent might make in negotiating an MDP or its purely stochastic precursor, the Markov Chain \cite{norris1998markov}. These included properties such as reachability of given sets of states, hitting times and probabilities of such sets of states, and other properties that are illuminated using distributions of program traces. The idea of reward is not really considered, except as one attached to a particular probabilistic program run across the MDP \cite{holzl2017markov}, which does not resemble the general reward that is typically considered with MDP optimisation problems.\

H\"olzl models finite MDPs using a type variable \texttt{'s} to represent the states of the MDP. He represents actions purely by their result, via a probability distribution over the states it might take an agent to. To achieve this, he uses the type \texttt{'s pmf} \cite{holzl2015formalized}, which represents probability mass functions (PMFs) over the type variable \texttt{'s}. A PMF is a function over countable states (in this instance) which returns the probability of a transition to that state. H\"olzl's formalisation is:\

\begin{lstlisting}[basicstyle=\footnotesize\ttfamily, mathescape = true]{}
locale Finite_Markov_Decision_Process = Markov_Decision_Process 
  K for K :: "'s $\Rightarrow$ 's pmf set" + fixes S :: "'s set"
  assumes S_not_empty: "S $\neq$ {}"
  assumes S_finite: "finite S"
  assumes K_closed: "$\bigwedge$s. s $\in$ S $\Longrightarrow$ ($\bigcup$D$\in$K s. set_pmf D) $\subseteq$ S"
  assumes K_finite: "$\bigwedge$s. s $\in$ S $\Longrightarrow$ finite (K s)"
\end{lstlisting}

\noindent where the following applies:\

\begin{itemize}
\item{\texttt{set\_pmf} here is a function which returns a set of all of the results of a probability mass function which have non-zero probability.}
\item{H\"olzl does not assume that the variable type \texttt{'s} is finite because he is inheriting part of the locale definition from another one where there are possibly infinite states.}
\item{This means that the parameter \texttt{S}, assumed to be finite and not empty, is the set of \textit{valid} states considered to be part of the MDP. The finiteness of this set defines the finiteness of the MDP.}
\item{\texttt{K} is a function on states which returns a finite set of all valid actions that may be taken from the specified state. Note the assumptions that K returns a finite set only, and that any actions in \texttt{K s} (where \texttt{s $\in$ S}) only lead to other states in \texttt{S}.}
\end{itemize}

We also note that H\"olzl uses a generalisation of the idea of policy called ``configurations'' \cite{holzl2017markov}. These can model policies that vary over time in a way that is not required for our work: our primary goal is to provide a formalisation of MDPs suitable for modelling reward optimisation methods, which typically assume stationary policies, where the choice of action does not vary over time, as they are provably sufficient for optimality \cite{puterman2014markov}. Configurations can also model stationary policies as a special case, but the underlying definition remains unsuited for our model since the extra definition required for configurations is not needed for our approach and would only complicate our proofs.\

Clearly, there is no general reward function or discounting factor defined as part of H\"olzl's locale, as it is not concerned with questions of reward or optimisation (in the general case). 
Our work, by contrast, will model real-valued rewards (see the next section). This allows us to prove the necessary results in optimisation theory that lead into dynamic programming and reinforcement learning.

In addition, the use of the parameter \texttt{S} in the above locale to denote the set of valid states, which is then constrained via an explicit assumption of finiteness instead of using the Isabelle sort \texttt{finite} over the type, leads to clumsiness in further proofs. More specifically, since functions in Isabelle are total,  we must then always specify the behaviour of any function over valid as well as invalid states and account for this in proofs.\

Additionally, the locale makes extensive use of the Giry monad \cite{giry1982categorical}. Our position is that the model we are looking to build should be accessible and usable for those working in the reinforcement learning field using the mathematics that most have encountered previously. When it comes to probability, this means using transition matrices and stochastic matrices, which are by far the most common elements in texts on this material e.g.\ in Puterman's classic textbook \cite{puterman2014markov}. We also gain the capacity to prove many additional theorems which require matrices to be fully realised, as we will show later in section \ref{subs:matr}.\

This requires more preliminary work in finding suitable type representations for the matrices we will work with, and in establishing the matrix properties we will make use of, but the additional proofs that might otherwise be difficult to formalise using a Giry monadic approach are important: for example, our proofs that the expected value associated with a policy can be found using the inverse of $(I - \gamma T_\pi)$, or of $\epsilon$-optimality of the value iteration algorithm (see sections \ref{subsec:Loperator} and \ref{subsec:valueiter}).\

Consequently, while using the ideas in H\"olzl's formalisation as the inspiration for our own work, we decided to define our own model with some crucial, necessary differences to focus on the optimisation questions that interest us.\

\section{The fundamentals of our formalisation}
\label{sec:fundform}
In this section, we describe the basic concepts that underlie our approach and review their mechanisation in Isabelle. We begin by discussing the specific differences between our approach and that of H\"olzl before introducing our ``slice'' based approach to understanding agent paths through an MDP.

\subsection{Formalising finite Markov Reward Processes}

As stated, our model is built around a locale inspired by H\"olzl's work, but with some important changes. Our definition is as follows:\

\begin{lstlisting}[basicstyle=\footnotesize\ttfamily, mathescape = true]{}
locale Finite_Markov_Reward_Process = 
  fixes K :: "'s::{finite_discrete_topology} $\Rightarrow$ 's pmf set"
    and R :: "('s $\times$ 's $\times$ 's pmf) $\Rightarrow$ real"
    and $\gamma$ :: real
  assumes K_f: "$\forall$s. finite' (K s)"
    and gamma_range:"0 $\leq$ $\gamma$ $\wedge$ $\gamma$ $\leq$ 1"
\end{lstlisting}

\noindent where:\

\begin{itemize}
\item{Similarly to H\"olzl's \texttt{Finite\_Markov\_Decision\_Process} locale, we use the type variable \texttt{'s} used to represent states, although we assert it belongs to the \texttt{finite\_discrete\_topology} type class described below. 
\item We also use the \texttt{K} function that returns a set of valid actions from a state, which again we assume is finite and non-empty for all \texttt{s} -- the Isabelle predicate \texttt{finite'} asserts both finiteness and non-emptiness.}
\item{\texttt{R} is a function on a 3-tuple of two states and an action, which returns a real valued reward.}
\item{\texttt{$\gamma$} is a discounting factor such that \texttt{0 $\leq$ $\gamma$ $\leq$ 1}.}
\end{itemize}

When first building our formalisation, we tried to inherit directly from H\"olzl's \texttt{Finite\_Markov\_Decision\_Process} locale. However, as alluded to in the previous section, its use of \texttt{S} as a finite set to contain all valid states in the MDP, and the consequent theoretical presence of \textit{invalid} states, greatly complicated the proof process. Every theorem required assumptions that any state was a member of \texttt{S}, and functions on states needed definition on invalid states as well as valid ones. To overcome these issues, we asserted that our type variable belongs to the \texttt{finite} type class, circumventing the need to consider the possibility of invalid states while retaining the finiteness properties we require.\ 

Subsequently, when we began working with bounded continuous functions over the state type (discussed in detail in sub-section \ref{subsec:bcfs}), we needed to show that functions on states were continuous. As states are a finite, discrete type, the only way we can show functions on them are continuous is to equip them with the discrete topology sort, and later to assert they form a metric space in order to use the Banach fixed point theorem (see sub-section \ref{subsec:final} for more detail).\ 

It is possible in Isabelle to assert that a type \texttt{'t} belongs to multiple type classes. However, when proving that a type derived from a type variable is an instance of a type class, one cannot assert multiple type classes on the type variables; if that is necessary for the proof, it will fail. As we will need to prove that several types derived from our state type are Banach spaces, we combine the type classes \texttt{finite}, \texttt{discrete\_topology} and \texttt{metric\_space} into a single type class, \texttt{finite\_discrete\_topology}:

\begin{lstlisting}[basicstyle=\footnotesize\ttfamily, mathescape = true]{}
class finite_discrete_topology 
  = finite + discrete_topology + metric_space
\end{lstlisting}

Now that we have defined the MDP with rewards, we need to build representations that match the discounted $Q$ and $V$ values discussed previously. We formalise $Q$ values, because if we have the $Q$ value function for a policy $\pi$, we can always find the $V$ value function by assuming we choose action $\pi(s_0)$ in our initial state $s_0$.\ 

Recall that H\"olzl used configurations as a generalisation of policies. As we are not inheriting directly from his locale, we can simplify our definitions. We define \texttt{policies} as the set of all functions from the state to probability distributions on the state (recall that the latter is the type for actions in the current context), where the function only returns a member of the set of valid actions on any state:\

\begin{lstlisting}[basicstyle=\footnotesize\ttfamily, mathescape = true]{}
definition policies :: "('s $\Rightarrow$ 's pmf) set" where
"policies = {p. ($\forall$s. p s $\in$ K s)}"
\end{lstlisting}

\subsection{A slice based approach to understanding agent paths through an MDP}

In order to model value, we need a way of showing the expectation of rewards assigned over time. To do this, we need to mechanise the probability of various paths being followed by an agent pursuing a policy over time. H\"olzl's method was to build a probability measure over the paths an agent might take -- his primary concern being reachability and other properties of the states the agent might hit.\ 

\begin{figure}
\begin{center}
\includegraphics[scale=0.8]{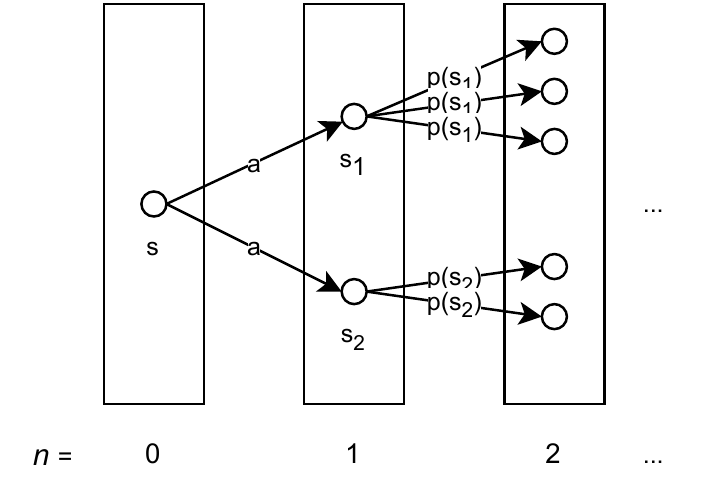}
\end{center}
\caption{An illustration of the slice approach. Each rectangle, or slice, at time step $n$ contains every possible state (e.g.\ $s_1$ and $s_2$ for slice 1) that an agent might be in by taking initial action $a$ then following policy $p$. The possible transitions from one time step to another are indicated by the labelled arrows (e.g.\ $p \ s_1$, $p \ s_2$ between step 1 and 2), with multiple arrows from a single action indicating that subsequent states are stochastically determined, with several possibilities.}
\label{fig:1}
\end{figure}

We need more concrete definitions for our purposes, so we take a different approach. We model the paths of an agent over finite time steps, and then consider the limit of those paths as the number of time steps approaches infinity. We introduce {\it slices} as a snapshot of an agent's possible path through an MDP, each slice capturing the possibilities at a given time step. For an illustration of this approach, see Figure \ref{fig:1}.\

We begin by defining the basics, namely a collection of functions that all take four parameters: the time step \texttt{n} being considered, the policy \texttt{p} that an agent is following, the initial state \texttt{s} that the agent is in, and the action \texttt{a} chosen in the first time step. After action \texttt{a} is performed, the agent always chooses action \texttt{p s'} for all subsequent states \texttt{s'}.\

\texttt{Qslice n p s a} returns the set of all the states that the agent might be in at time step $n$. In the definition below, \texttt{set\_pmf} is a function on an action which returns the set of all states that that action might choose with non-zero probability.\

\begin{lstlisting}[basicstyle=\footnotesize\ttfamily, mathescape = true]{}
fun Qslice :: "nat $\Rightarrow$ ('s $\Rightarrow$ 's pmf) $\Rightarrow$ 's $\Rightarrow$ 's pmf $\Rightarrow$ 's set" 
where
  "Qslice 0 p s a = {s}" 
| "Qslice (Suc 0) p s a = set_pmf a" 
| "Qslice (Suc (Suc n)) p s a 
    = ($\bigcup$s'$\in$Qslice (Suc n) p s a. set_pmf (p s'))"
\end{lstlisting}

\texttt{Qslicep n p s a s'} returns the total probability that, by any possible route given initial action \texttt{a} and subsequent policy \texttt{p}, an agent will be in state \texttt{s'} at time step \texttt{n}. We see in the definition below \texttt{pmf a s'} used as a function on an action and a state; it returns the probability of the state being picked by the action.\

\begin{lstlisting}[basicstyle=\footnotesize\ttfamily, mathescape = true]{}
fun Qslicep :: "nat $\Rightarrow$ ('s $\Rightarrow$ 's pmf) $\Rightarrow$ 's $\Rightarrow$ 's pmf $\Rightarrow$ 's $\Rightarrow$ real" 
where
  "Qslicep 0 p s a s' = (if (s' = s) then 1 else 0)"
| "Qslicep (Suc 0) p s a s' = pmf a s'"
| "Qslicep (Suc (Suc n)) p s a s' 
    = ($\sum$s''$\in$a. pmf a s'' * Qslicep (Suc n) p s'' (p s'') s')"
\end{lstlisting}

\texttt{RQslice n p s a} returns the reward expected by an agent at time step \texttt{n} for taking a single action and transitioning to a new state at time step \texttt{n+1}. We do not take discounting into account when calculating \texttt{RQslice} -- it is the expected reward earned, not the value we place on it. Note that it is a single step's worth of reward only.\

\begin{lstlisting}[basicstyle=\footnotesize\ttfamily, mathescape = true]{}
fun RQslice :: "nat $\Rightarrow$ ('s $\Rightarrow$ 's pmf) $\Rightarrow$ 's $\Rightarrow$ 's pmf $\Rightarrow$ real" 
  where
  "RQslice 0 p s a = ($\sum$s'$\in$a. (pmf a s') * R (s, s', a))" 
| "RQslice (Suc n) p s a =  
    ($\sum$s'$\in$Qslice (Suc n) p s a. Qslicep (Suc n) p s a s' * 
    ($\sum$s''$\in$p s'. (pmf (p s') s'') * R (s', s'', (p s'))))"
\end{lstlisting}

To ensure that these are consistent and represent our expectations for MDPs, we prove some basic properties of these functions. We begin by showing that for each function the recursive definition can be evaluated in different ways:\

\noindent \begin{minipage}{\linewidth}
\begin{lstlisting}[basicstyle=\footnotesize\ttfamily, mathescape = true]{}
lemma Qslice_rec:
  fixes p :: "'s $\Rightarrow$ 's pmf" and s :: 's and a :: "'s pmf"
  shows "Qslice (Suc m) p s a 
    = ($\bigcup$s'$\in$a. Qslice m p s' (p s'))"
    
lemma Qslicep_rec:
  fixes s' s''' :: 's and a' :: "'s pmf" and p :: "('s $\Rightarrow$ 's pmf)"
  assumes "a'$\in$K s'"
  shows "Qslicep (Suc m) p s' a' s''' 
    = ($\sum$s''$\in$a'. pmf a' s'' * Qslicep m p s'' (p s'') s''')"
    
lemma Qslicep_rec2:
  fixes s s' :: 's and a :: "'s pmf" and p :: "('s $\Rightarrow$ 's pmf)"
  assumes a_in_Ks:"a$\in$K s" and p_in_pol:"p$\in$policies"
  shows "Qslicep (Suc (Suc m)) p s a s' 
    = ($\sum$s''$\in$Qslice (Suc m) p s a. 
      Qslicep (Suc m) p s a s'' * pmf (p s'') s')"

lemma RQslice_rec:
  fixes s s' :: 's and a :: "'s pmf" and p :: "('s $\Rightarrow$ 's pmf)"
  assumes a_in_Ks:"a$\in$K s" and p_in_pol:"p$\in$policies"
  shows "RQslice (Suc n) p s a 
    = ($\sum$s'$\in$a. (pmf a s') * RQslice n p s' (p s'))"
\end{lstlisting}
\end{minipage}

These lemmas are useful for proving the theorems we will go on to establish as they show the equivalence of several different definitions for the functions.\

The biggest difficulty we had proving these and later theorems over the slice functions arose when showing that it is possible to evaluate various values across all the states in a slice at time step \texttt{n} and those in the slice at the subsequent time step \texttt{n+1}, and produce identical results.\

Several different lemmas of this type were required, and despite their similarity, it was not evident how we might formalise a general result; instead we proved each separately. As an example, we showed that:\

\begin{lstlisting}[basicstyle=\footnotesize\ttfamily, mathescape = true]{}
lemma Qslice_pol_summation2:
  fixes s s'' :: 's and a :: "'s pmf" and p :: "('s $\Rightarrow$ 's pmf)" 
  assumes a_in_Ks:"a$\in$K s" and p_in_pol:"p$\in$policies" 
    and m_gr_0:"m > 0" and s''_in_a:"s''$\in$a"
  shows "($\sum$s'''$\in$Qslice m p s'' (p s''). 
    pmf a s'' * Qslicep m p s'' (p s'') s''' * pmf (p s''') s') 
      = ($\sum$s'''$\in$Qslice (Suc m) p s a. 
        pmf a s'' * Qslicep m p s'' (p s'') s''' 
          * pmf (p s''') s')"
\end{lstlisting}

This specific lemma shows that summing the total probability of being in a given state at a given time step of a particular set of paths is the same regardless of whether we measure the paths from an initial state $s$ or from some subsequent state $s'$ after accounting for the transition between $s$ and it.\

We next show that \texttt{Qslicep} produces a probability distribution over the states at any time step \texttt{n}:

\noindent \begin{minipage}{\linewidth}
\begin{lstlisting}[basicstyle=\footnotesize\ttfamily, mathescape = true]{}
lemma Qslicep_nn: "Qslicep m p s a s' $\geq$ 0"

lemma Qslicep_1:
 fixes s :: 's and a :: "'s pmf" and p :: "('s $\Rightarrow$ 's pmf)" 
   and m :: nat
 assumes a_in_Ks:"a$\in$K s" and p_in_pol:"p$\in$policies"
 shows "($\sum$s'$\in$Qslice (Suc m) p s a. Qslicep (Suc m) p s a s') = 1"  
\end{lstlisting}
\end{minipage}

\noindent and that \texttt{s' $\in$ Qslice n p s a} if and only if \texttt{Qslicep n p s a s' > 0}:

\begin{lstlisting}[basicstyle=\footnotesize\ttfamily, mathescape = true]{}
lemma Qslice_Qslicep: 
  fixes s s' :: 's and p :: "('s $\Rightarrow$ 's pmf)" and a :: "'s pmf"
  assumes a_in_Ks:"a$\in$K s" and p_in_pol:"p$\in$policies"
  shows "s' $\in$ Qslice m p s a = (Qslicep m p s a s' > 0)"
\end{lstlisting}

\subsection{Proving convergence properties}
\label{subsec:expected}
These basic properties proven, we go on to build our definition of value on our MDP model. We do this using two functions, as described next.\

\texttt{Qexpectedn n p s a} gives us the expected value up to time step \texttt{n} of an agent beginning in state \texttt{s}, taking action \texttt{a} and subsequently following policy \texttt{p}. The discount factor \texttt{$\gamma$} is taken into account in this definition.\

\begin{lstlisting}[basicstyle=\footnotesize\ttfamily, mathescape = true]{}
definition 
 Qexpectedn :: "nat $\Rightarrow$ ('s $\Rightarrow$ 's pmf) $\Rightarrow$ 's $\Rightarrow$ 's pmf $\Rightarrow$ real"
  where
"Qexpectedn n p s a = ($\sum$i=0..n. $\gamma$^i * RQslice i p s a)"
\end{lstlisting}
\noindent where \texttt{$\gamma$\string^i} indicates $\gamma^i$.

\texttt{Qexpected p s a} takes the limit of the \texttt{Qexpectedn} function as \texttt{n} approaches infinity. We will prove that this function represents the $Q$ value as used in MDP literature (see section \ref{Qmathdef}) by showing that we can derive the Bellman equation using it in section \ref{eqn:bellman}.

\begin{lstlisting}[basicstyle=\footnotesize\ttfamily, mathescape = true]{}
definition Qexpected :: "('s $\Rightarrow$ 's pmf) $\Rightarrow$ 's $\Rightarrow$ 's pmf $\Rightarrow$ real" 
where
"Qexpected p s a = lim ($\lambda$n. Qexpectedn n p s a)"
\end{lstlisting}

We define a terminal state in our representation as one where all possible actions produce a reward of zero and do not transition the agent into a different state. In other words, if an agent is in a terminal state, neither total reward earned nor the state of the agent will change in subsequent slices.\

\begin{lstlisting}[basicstyle=\footnotesize\ttfamily, mathescape = true]{}
definition terminal_state :: "'s $\Rightarrow$ bool" where
  "terminal_state s = 
      ($\forall$s' a'. R (s, s', a') = 0 $\wedge$ ($\bigcup$a$\in$K s. set_pmf a) = {s})"
\end{lstlisting}

We say that an MDP has inevitable terminal states under a given policy \texttt{p} if and only if for any agent pursuing that policy, after some finite time, the slice of possible states of the agent only includes terminal states:

\begin{lstlisting}[basicstyle=\footnotesize\ttfamily, mathescape = true]{}
definition inevitable_term_state :: "('s $\Rightarrow$ 's pmf) $\Rightarrow$ bool" where
  "inevitable_term_state p = 
      ($\forall$s. $\forall$a$\in$K s. $\exists$n. $\forall$s'$\in$Qslice n p s a. terminal_state s')"
\end{lstlisting}

We prove that \texttt{Qexpectedn n p s a} converges as \texttt{n $\to \infty$} next, and thus \texttt{Qexpected p s a} has a definite value, under either of two different assumptions:\

\begin{itemize}
    \item When \texttt{inevitable\_term\_state} is true under policy \texttt{p}. 
    \item When \texttt{$\gamma$<1}. This second assumption is the one that we will use in our later proofs throughout section \ref{sec:proofopt}, but specifically in subsection \ref{subsec:final}, as it is fundamental, in the context of a finite MDP, to proving an optimal policy exists. 
\end{itemize}

The proof under the first assumption is essentially trivial, as knowing that the agent can only be in some terminal state at some finite time step means we simply must sum our rewards until the point where \texttt{Qslice n p s a} is a set of only terminal states. As there exists an \texttt{n} where this must be true, convergence is guaranteed. The Isabelle theorem is thus:\

\begin{lstlisting}[basicstyle=\footnotesize\ttfamily, mathescape = true]{}
theorem convergence_Qexpectedn_term_state:
  assumes "inevitable_term_state p" and "a$\in$K s" 
      and "p$\in$policies"
    shows "convergent ($\lambda$n. Qexpectedn n p s a)"
\end{lstlisting}

The proof under the second assumption is more complicated and depends on the reward function \texttt{R} being bounded (above and below) by some \texttt{M}, which as it is acting on a finite set of states and actions, it must be. That established, our discounted \texttt{RQslice} function must be less than \texttt{$\sum_{i=0}^n\gamma^i M$} for any $n$. As \texttt{$\gamma$<1}, this is a geometric series, so we know it converges. By the usual comparison and absolute value series tests \cite{ayres2013schaum,fleuriot2000}, we know that our \texttt{Qexpectedn} value converges too as \texttt{n} approaches infinity. The resulting Isabelle theorem is as follows:\

\begin{lstlisting}[basicstyle=\footnotesize\ttfamily, mathescape = true]{}
theorem convergence_Qexpectedn_discount:
  assumes "a$\in$K s" and "p$\in$policies" and "$\gamma$ < 1"
  shows "convergent ($\lambda$n. Qexpectedn n p s a)"
\end{lstlisting}

\subsection{Deriving the Bellman equation}

Our final step here is to show that our method of calculating the $Q$ value of a given state-action-policy combination matches the Bellman equation (see section \ref{sec:Background}) \cite{barron1989bellman}.\

We do this by first showing that it is possible to evaluate \texttt{Qexpectedn} recursively:\

\begin{lstlisting}[basicstyle=\footnotesize\ttfamily, mathescape = true]{}
lemma Qexpectedn_rec:  
  fixes s s' :: 's and a :: "'s pmf" and p :: "('s $\Rightarrow$ 's pmf)"
  assumes a_in_Ks:"a$\in$K s" and p_in_pol:"p$\in$policies"
  shows "Qexpectedn (Suc n) p s a = 
    ($\sum$s'$\in$a. (pmf a s') * (R (s, s', a) + 
        $\gamma$ * Qexpectedn n p s' (p s')))"
\end{lstlisting}

We then use this to show that it is possible to evaluate \texttt{Qexpectedn} by ``splitting'' the value as such:\

\begin{lstlisting}[basicstyle=\footnotesize\ttfamily, mathescape = true]{}
lemma Qexpectedn_split:  
  fixes s :: 's and a :: "'s pmf" and p :: "('s $\Rightarrow$ 's pmf)"
  assumes a_in_Ks:"a$\in$K s" and p_in_pol:"p$\in$policies"
  shows "Qexpectedn (Suc n+m) p s a = Qexpectedn n p s a + 
    ($\sum$s'$\in$Qslice (Suc n) p s a. $\gamma$^(Suc n) *
         Qslicep (Suc n) p s a s' * Qexpectedn m p s' (p s'))"
\end{lstlisting}

This (along with a few minor lemmas showing properties of limits) in turn leads to the proof that \texttt{Qexpected p s a} produces the same result as the Bellman equation:\

\label{eqn:bellman}
\begin{lstlisting}[basicstyle=\footnotesize\ttfamily, mathescape = true]{}
theorem Qexpected_split_Bellman:
  fixes a :: "'s pmf" and s :: 's and p :: "'s $\Rightarrow$ 's pmf"
  assumes "a$\in$K s" and "p$\in$policies" 
    and "$\wedge$s a'. a'$\in$K s $\Longrightarrow$ convergent ($\lambda$n. Qexpectedn n p s a')"
  shows "Qexpected p s a = ($\sum$s'$\in$a. (pmf a s') * (R (s, s', a) +
                              $\gamma$ * Qexpected p s' (p s')))"
\end{lstlisting}

Note that for this proof, we assume convergence of \texttt{Qexpectedn} rather than any specific assumption which proves that convergence. From this general proof, we can derive the Bellman equation under the assumption that \texttt{$\gamma$<1} or \texttt{inevitable\_terminal\_states p}.\

We then define \texttt{Vexpected p s = Qexpected p s (p s)}; in other words, the V value is simply the Q value where we replace the chosen action by that dictated by our choice of policy.\

We have now shown that we can derive Bellman's equation in full from our locale assumptions and the model of value we built based on a na\"ive understanding of an agent gathering reward over time. This (along with our other results) gives us confidence that our model is not flawed and accurately represents a Markov decision process.\

\section{Proving the existence of an optimal policy}
\label{sec:proofopt}

In this section, we cover the non-trivial mechanisation of the proof of the existence of an optimal policy on all finite MDPs, if a discounting factor exists which is less than one.

\subsection{Why is an optimal policy important?}

The purpose of using an MDP is typically to come up with the ideal strategy for optimising some discretisable process: to find some policy that, regardless of an agent's starting state or choice of action, we can expect to produce the highest possible total reward if pursued. As noted before (see section \ref{sec:Background}), we usually evaluate this potential total reward using the concept of a discounted value. We call this policy an optimal policy.\

A very important consideration, then, is does such an optimal policy exist? Clearly, given there are finite states and finite actions available on those states, our choice of policies is finite too. For any individual choice of initial state-action combination, there will be a policy that produces the highest value when followed subsequently. But the question of the existence of a policy that produces a maximal result for \textit{any} choice of initial state and action is critical to knowing that an MDP can fulfil its purpose.\

We use Puterman's proof of the existence of an optimal policy \cite{puterman2014markov} as the basis for our own formal proof. It is important to note though that Puterman approaches this question from a more general starting point -- his proof is adequate for finite or infinite MDPs (using slightly different methods at various points), and considers non-stationary policies (policies whose choice function might vary over time) before demonstrating a stationary policy is adequate. We begin by assuming that all policies are stationary and that our MDP is finite -- as previously noted, stationary policies are sufficient to find optimality (as we will establish here), and the scope of our formalisation is over finite MDPs.\

Note that we are skipping the detail of the individual proofs themselves, instead giving the outline of the structure of Puterman's proof so that it is clear what is needed from the description of our formal version. Moreover, in many cases when discussing the mathematical background, we will state that we mechanise particular statements without providing the formal proof, except as a sketch at most. The interested reader can find details of the pen-and-paper proof in chapters 5 and 6 and appendix C of Puterman's book, and can consult our Isabelle theories for additional details pertaining to the formalisation.\

As we review the various elements of the proof, we will begin by going over the mathematics and then discuss the details of the formalisation.\

\subsection{Functions on states as a vector space}
\label{subsec:bcfs}

First, note that all real functions on our states are over a finite set and are therefore bounded in magnitude. Consider them as belonging to the vector space $V$ of bounded real-valued functions on the states. In $V$ each component of a vector represents the result of the function on a particular state. Note that $V$ is closed under addition and scalar multiplication as expected \cite{kreyszig1978introductory}. The dimension of the vector space is equal to the number of states that we have.\

Equip $V$ with the supremum norm, which is the supremum in the reals of the norms of its components (recall each component is $v(s)$ for some state $s$). Each supremum is attained by each $v\in V$ as we have finite states. The definition is then as follows:\ 

\begin{align}
\forall v \in V. \lVert v\rVert &= \sup_{s\in S} \lvert v(s) \rvert = \max_{s \in S} \lvert v(s) \rvert
\end{align}

Then equip $V$ with a partial order where a vector $v$ is less than or equal to another $v'$ if and only for all states $s$, $v(s)\leq v'(s)$. Because this is a partial order, we have no guarantee that arbitrary vectors in $V$ are able to be compared using them -- if there exists $s,s'$ such that $v(s)\leq v'(s)$ and $v'(s')\leq v(s')$, then $v$ and $v'$ cannot be ordered with respect to each other. This partial ordering is given by:\

\begin{align}
\forall v, v' \in V. v\leq v' &\iff \forall s. v(s) \leq v'(s)
\end{align}

A \textit{Cauchy sequence} in $V$ is a sequence $v_n$ of vectors where $$\forall \epsilon>0. \exists N. \forall n,m \geq N. \lVert v_m - v_n\rVert < \epsilon$$

\noindent and a \textit{complete space} is one in which the limit of all Cauchy sequences in the space is attained within the space \cite{kreyszig1978introductory}. A \textit{Banach space} is simply a complete, normed vector space, and thus we know that $V$ is a Banach space: this will be critical to the later proof in section \ref{subsec:final} when we make use of the Banach fixed point theorem, a property of Banach spaces and certain functions (contraction mappings) over them.\

Isabelle has several formalisations of vectors but none that exactly matches what we need. There is an often-used vector type, \texttt{vec}, within the ``Finite Cartesian Product'' theory of the main Isabelle/HOL analysis library, and it would allow us to define vectors from our real functions over the finite state type. Unfortunately, it lacks the supremum norm required by Puterman's proof and has a Euclidean norm instead.\

However, Isabelle's analysis library has a formal theory of bounded continuous functions, with a type of the same name which has a supremum norm defined over it. This type also belongs to the normed vector space type class and the complete space type class, so we can perform addition and scalar multiplication over it and we know it is a Banach space.\

In this theory, \texttt{('a, 'b) bcontfun} is a type that represents bounded continuous functions from some type variable \texttt{'a} (which must be an instance of the \texttt{topological\_space} type class) to the type variable \texttt{'b} (which must be an instance of the \texttt{metric\_space} type class).\

If we equip our state space with the discrete topology (where any subset is both an open and closed set), any function on it is continuous through the topological definition of continuity (where a function is continuous if the preimage of any open set is open), and the type variable we use to represent it would be an instance of the \texttt{topological\_space} type class. As the state space is finite, we also know that functions on it must be bounded. So we can represent real functions over a finite state space as bounded continuous functions this way, with all the properties of a Banach space and the supremum norm we are looking for.\

We still have to define a partial ordering over the bounded continuous functions, which we do. Note that \texttt{apply\_bcontfun} accesses the underlying function from a \texttt{bcontfun} and obtains its result.\

\noindent \begin{minipage}{\linewidth}
\begin{lstlisting}[basicstyle=\footnotesize\ttfamily, mathescape = true]{}
definition less_eq_bcontfun :: 
  "('a, 'b) bcontfun $\Rightarrow$ ('a, 'b) bcontfun $\Rightarrow$ bool" 
  where "less_eq_bcontfun a b 
    = ($\forall$i. apply_bcontfun a i $\leq$ apply_bcontfun b i)"

definition less_bcontfun :: 
  "('a, 'b) bcontfun $\Rightarrow$ ('a, 'b) bcontfun $\Rightarrow$ bool" 
  where "less_bcontfun a b = (a $\leq$ b $\wedge$ $\lnot$ b $\leq$ a)"
\end{lstlisting}
\end{minipage}

\subsection{Operators over functions on states as matrices}
\label{subs:matr}

Next, we return to laying the groundwork for Puterman's proof: we consider linear operators over $V$ as matrices in their own vector space $W$, with the usual understanding of matrix-vector multiplication representing its operation. As normal, a matrix $A$ may be shown as invertible with an inverse $A^{-1}$, if $AA^{-1} = A^{-1}A = I$, where $I$ is the identity matrix \cite{lipschutz2018schaum}.\

For these operators, we define their norm as the operator norm \cite{beauzamy1988introduction} against the result of their operation on vectors in $V$. For a matrix $A$, this is the least upper bound on $\frac {\lVert Av \rVert}{\lVert v \rVert}$, meaning the greatest scaling we expect its operation to have on the norm of any vector $v$. It can be shown that the operator norm for $A$ can be found by taking the supremum of the norm of any vectors of at most norm 1 after they are multiplied by $A$ \cite{beauzamy1988introduction}:\

\begin{equation}
\begin{aligned}
\forall A \in W . \lVert A \rVert &= \inf\{c\geq0:\forall v \in V: \lVert Av \rVert \leq c\lVert v\rVert\}\\
&= \sup\{\lVert{Av}\rVert : \lVert v \rVert \leq 1, v\in V\}
\end{aligned}
\end{equation}

In this case, as we are using the supremum norm for vector space $V$, we can show that the operator norm is equal to the highest sum of absolute values in any row of $A$. If we take $A_{i,j}$ as being the component in row $i$ and column $j$ of the matrix $A$, then:\

\begin{align}
\forall A \in W : \lVert A \rVert &= \sup_i\{\sum_j \lvert A_{i,j} \rvert\}
\end{align}

In the Isabelle analysis library's ``Finite Cartesian Product'' theory, matrices are represented as vectors-of-vectors (abbreviated at \texttt{mat}). Again, unfortunately, the norm definition does not match what we need (the Euclidean norm is used), and in any case the vector type we are using is not defined for multiplication by members of the \texttt{mat} type.\

In order to match Puterman's usage of matrices, we have two main requirements:

\begin{enumerate}
    \item The matrix type should be equipped with the operator norm.
    \item It should provably be a complete space.
\end{enumerate}

\noindent Unfortunately, no matrix type in Isabelle's current libraries has exactly what we need. However, the \texttt{sq\_mtx} type of square matrices in the ``SQ MTX'' theory, part of the ``Matrices for ODEs'' entry in the Archive of Formal Proof \cite{y2020matrices} does have a notion of operator norm, which we can adapt to meet our requirements. Note that if \texttt{A} is of this type, we use the notation \texttt{A\mbox{\textdollar\textdollar}i\mbox{\textdollar}j} to refer to its component in row \texttt{i} and column \texttt{j}. This \texttt{sq\_mtx} type is restricted to representing real-valued square matrices, but that is all we need for our work.\

The \texttt{sq\_mtx} type has multiplication defined against the \texttt{vec} type, not the bounded continuous function type that we are using. In order to use this definition for matrix multiplication and the various theorems that prove its properties, we demonstrate that a bijection exists between \texttt{vec} (which does have matrix multiplication defined against it) and bounded continuous functions.\ 

Vectors, represented as \texttt{('a, b') vec} type, can be defined in Isabelle using Cartesian products, where \texttt{'a} gives the type of the vector components and \texttt{'b} is a finite type variable whose cardinality represents the dimension of the vector. The second type, by virtue of its finiteness, provides an index into the vector, enabling us to refer to its components. Thus, a \texttt{('a, b') vec} for a type variable \texttt{'b} is essentially the function space \texttt{'a} $\Rightarrow$ \texttt{'b} \cite{harrison-euclidean}. For a vector \texttt{v} of type \texttt{vec}, the notation \texttt{v\mbox{\textdollar}i} then represents its \texttt{i}$^{th}$ component and the notation \texttt{($\chi$ i. f i)} (where \texttt{f i} can be any function) defines a vector whose \texttt{i}$^{th}$ component is equal to \texttt{f i}. 

A similar notation, namely \texttt{('a, 'b) bcontfun} is used for bounded continuous functions, although, in this case, this type refers to the underlying function space \texttt{'a::topological\_space} $\Rightarrow$ \texttt{'b::metric\_space} as previously noted. 

Given these two representations of vectors and bounded continuous functions, we define the following bijections between them:\

\begin{lstlisting}[basicstyle=\footnotesize\ttfamily, mathescape = true]{}
definition bcontfun_of_vec :: 
 "('a::metric_space, 
   'b::finite_discrete_topology) vec $\Rightarrow$ ('b, 'a) bcontfun" where
"bcontfun_of_vec v = Bcontfun($\lambda$i. v$\mbox{\textdollar}$i)"


definition vec_of_bcontfun :: 
 "('b::finite_discrete_topology, 
   'a::metric_space) bcontfun $\Rightarrow$ ('a, 'b) vec" where
"vec_of_bcontfun f = ($\chi$ i. (apply_bcontfun f) i)"

lemma bcf_inv[simp]:
  shows "vec_of_bcontfun (bcontfun_of_vec v) = v" 
  and "bcontfun_of_vec (vec_of_bcontfun f) = f"
\end{lstlisting}

We define a new matrix multiplication operation over a bounded continuous function by first casting it to a vector of the \texttt{vec} type, then perform matrix-vector multiplication, and finally casting the result back to a bounded continuous function. This method also ensures that the existing body of proofs regarding matrix-vector multiplication are available to us when we need to prove the properties of our matrix-bounded continuous function multiplication, greatly easing the amount of work we have to do here (note the \texttt{sq\_mtx\_vec\_mult} function simply performs a matrix-vector multiplication):\

\begin{lstlisting}[basicstyle=\footnotesize\ttfamily, mathescape = true]{}
definition sq_mtx_bcf_mult ::
  "('m::finite_discrete_topology) sq_mtx $\Rightarrow$ 
    ('m, real) bcontfun $\Rightarrow$ ('m, real) bcontfun" (infixl "$*_M$" 90) 
where
"A $*_M$ x =
    bcontfun_of_vec (sq_mtx_vec_mult A (vec_of_bcontfun x))"
\end{lstlisting}

Note that we use the notation \texttt {r $*_R$ A} to represent scalar multiplication of a matrix \texttt{A} by scalar \texttt{r} and the notation \texttt {A $*_M$ v} to represent matrix multiplication of a bounded continuous function \texttt{v} by a matrix \texttt{A}.\

We go on to prove a wide variety of typical linear algebra results of matrix-vector multiplication are still true against our bounded continuous function definition. The proofs for these fall out quickly from our use of the existing matrix-vector multiplication in our new definition as we have established a bijection between the two types previously. Some example theorems are the following:\

\begin{lstlisting}[basicstyle=\footnotesize\ttfamily, mathescape = true]{}
lemma mtx_bcf_mult_add_rdistr: "(A + B) $*_M$ x = A $*_M$ x + B $*_M$ x"
  unfolding sq_mtx_bcf_mult_def 
  by (simp add: bcontfun_vec_add mtx_vec_mult_add_rdistr)
  
lemma sq_mtx_times_bcf_assoc: "(A * B) $*_M$ x = A $*_M$ (B $*_M$ x)"
  by (metis bcf_inv(1) sq_mtx_bcf_mult_def sq_mtx_times_vec_assoc)
\end{lstlisting}

We then equip these matrices with the operator norm against this function on bounded continuous functions. Our definition here performs a matrix multiplication on a bounded continuous function, which gives us a vector that we then cast as a bounded continuous function again. This means that the \texttt{onorm} function, also denoted by $\lVert$\_$\rVert_{om}$, measures the change in the norm on the bounded continuous function using its supremum norm, which is the behaviour we need.

\begin{lstlisting}[basicstyle=\footnotesize\ttfamily, mathescape = true]{}
abbreviation om_norm :: 
  "(('a::{metric_space, real_normed_vector, semiring_1}, 
    'b::finite_discrete_topology) vec, 'b) vec $\Rightarrow$ real"
  where "$\lVert$A$\rVert_{om}$ $\equiv$ onorm ($\lambda$x. bcontfun_of_vec (bcf_mtx_mult x A))"
\end{lstlisting}

This is then used to define the norm in our proof that our square matrix type forms a normed vector space:

\begin{lstlisting}[basicstyle=\footnotesize\ttfamily, mathescape = true]{}
definition norm_sq_mtx :: "'a sq_mtx $\Rightarrow$ real" 
  where "$\lVert$A$\rVert$ = $\lVert$to_vec A$\rVert_{om}$"
\end{lstlisting}

We go on to prove its properties as required, beginning with verifying that the operator norm thus defined is equal to the highest sum of the absolute values of the elements of any row in the matrix:\

\begin{lstlisting}[basicstyle=\footnotesize\ttfamily, mathescape = true]{}
theorem sq_mtx_norm_value:
  fixes A :: "('b::finite_discrete_topology) sq_mtx"
  shows "$\lVert$A$\rVert$ = Max ($\bigcup$i. {($\sum$j. norm (A$\mbox{\textdollar\textdollar}$i$\mbox{\textdollar}$j))})"
\end{lstlisting}  

\noindent The next result: 

\begin{lstlisting}[basicstyle=\footnotesize\ttfamily, mathescape = true]{}  
theorem sq_mtx_norm_mult:
  fixes A B :: "('b::finite_discrete_topology) sq_mtx"
  shows "$\lVert$A * B$\rVert$ $\leq$ $\lVert$A$\rVert$ * $\lVert$B$\rVert$" 
\end{lstlisting}
\noindent is also important to many subsequent proofs, in particular when dealing with stochastic matrices.

The \texttt{sq\_mtx} type was originally shown to form a complete space, but we reprove this using our new operator norm, establishing that we still have a Banach space.\

\subsection{Stochastic matrices}

When using matrix operators, we will usually be dealing with matrices that capture the probability transitions of a particular policy $\pi$, denoted $T^\pi$. Each component $T^\pi_{i,j}$ represents the probability of transitioning to state $j$ under action $\pi(i)$. As these represent transitions, we refer to them as transition matrices, but they are more generally examples of stochastic matrices \cite{chen2021introduction}. As each row of a matrix represents a probability distribution, it should be clear that each row therefore sums to 1, with each element being non-negative.\

We establish the notion of a stochastic matrix in Isabelle via a predicate which verifies this property:

\begin{lstlisting}[basicstyle=\footnotesize\ttfamily, mathescape = true]{}
definition stochastic_mtx :: 
  "'b::finite_discrete_topology sq_mtx $\Rightarrow$ bool" where
"stochastic_mtx A = (($\forall$i. ($\sum$j. A$\mbox{\textdollar\textdollar}$i$\mbox{\textdollar}$j) = 1)  $\wedge$ ($\forall$i j. A$\mbox{\textdollar\textdollar}$i$\mbox{\textdollar}$j $\geq$ 0))"
\end{lstlisting}

\noindent and then prove some of elementary properties of stochastic matrices, starting with the fact that the identity matrix is a stochastic matrix:

\begin{lstlisting}[basicstyle=\footnotesize\ttfamily, mathescape = true]{}
theorem stochastic_mtx_1:
  shows "stochastic_mtx 1"
\end{lstlisting}

Next, we prove that the product of any two stochastic matrices is also a stochastic matrix and likewise for a exponentiation of a stochastic matrix. We inherit matrix exponentiation and its properties for the \texttt{sq\_mtx} type from the fact that it is proven as an instance of the \texttt{real\_normed\_algebra\_1} type class, which has exponentiation defined against it.

\begin{lstlisting}[basicstyle=\footnotesize\ttfamily, mathescape = true]{}
theorem stochastic_mtx_mult:
  fixes A B :: "'b::finite_discrete_topology sq_mtx"
  assumes "stochastic_mtx A" "stochastic_mtx B"
  shows "stochastic_mtx (A * B)"

theorem mat_pow_stochastic_mtx:
  fixes A :: "'b::finite_discrete_topology sq_mtx" and n :: nat
  assumes "stochastic_mtx A"
  shows "stochastic_mtx (A^n)"
\end{lstlisting}

\noindent Note that \texttt{A\^}\texttt{n} represents a matrix \texttt{A} raised to the power \texttt{n}. We also show that the operator norm is 1 for any stochastic matrix (which falls very simply out of our definition of a stochastic matrix):\

\noindent \begin{minipage}{\linewidth}
\begin{lstlisting}[basicstyle=\footnotesize\ttfamily, mathescape = true]{}
theorem stochastic_mtx_norm:
  fixes A :: "'b::finite_discrete_topology sq_mtx"
  assumes "stochastic_mtx A"
  shows "$\lVert$A$\rVert$ = 1"
\end{lstlisting}
\end{minipage}

\subsection{Spectral radius using Gelfand's formula}
\label{subsec:gelfform}

We now build a formal notion of spectral radius on our matrices. We denote the spectral radius of a matrix $A$ by $\rho(A)$. Normally the spectral radius of a matrix is defined as the largest of its eigenvalues \cite{lipschutz2018schaum} -- an eigenvalue being the scalar by which $A$ multiplies a vector that otherwise is unchanged by matrix-vector multiplication.\

We will not use this approach though since our purpose here is to further our formalisation of MDPs and the notion of spectral radius only plays a small part in Puterman's proof, namely to establish the invertibility of certain matrices, for which we only need Gelfand's formula: $\rho(A) = \lim_{n\to\infty} \lVert A^n \rVert^{1/n}$ \cite{gelfand1941normierte}. This is especially important for proving that we can express the expected value vector of a policy in terms of a relatively simple inverse matrix (\ref{eqn:inversevpi}), and for deriving some other results in section \ref{subsec:Loperator}. Using this, we then need to show that:\

\begin{align}
\label{eqn:specradinv}
    \rho(I - A)<1\implies \exists A^{-1}. A^{-1} = \lim_{N\to\infty} \sum_{n=0}^N (I-A)^n
\end{align}

Therefore, we express the spectral radius using the notion of matrix exponentiation to match Gelfand's formula. Thankfully, the latter is already defined for types satisfying the \texttt{real\_normed\_algebra\_1} sort, within a theory belonging to the ``Ergodic Theory'' session in the AFP \cite{Ergodic_Theory-AFP}:\

\begin{lstlisting}[basicstyle=\footnotesize\ttfamily, mathescape = true]{}
definition spectral_radius::"'a::real_normed_algebra_1 $\Rightarrow$ real"
  where "spectral_radius x = Inf {root n (norm(x^n))| n. n>0}"
\end{lstlisting}

\noindent where \texttt{root n x} denotes the $n^{th}$ root of real number $x$ and \texttt{Inf} denotes the infimum operator. This definition is easily shown to be equivalent to the limit one from Gelfand's formula and then used to formalise statement (\ref{eqn:specradinv}) as follows: 

\begin{lstlisting}[basicstyle=\footnotesize\ttfamily, mathescape = true]{}
theorem specrad_inverse_matrix:
  fixes A :: "'b::finite_discrete_topology sq_mtx" and L :: real
  assumes "spectral_radius (1-A) < 1"
  shows "mtx_invertible A" "A$^{-1}$ = lim ($\lambda$N. $\sum$n=0..N. (1 - A)^n)"
\end{lstlisting}

The formal proof of the above result is lengthy and depends on the fact that we have shown that the matrix types we are using form a Banach space. This means we can find a limit to the series, that we then show is the inverse of the matrix $A$. 

Even though formalising this result involved a fair amount of work, we will not be going into the details here for the sake of brevity (and to avoid distracting from the main thrust of the paper). We refer the interested reader to the associated Isabelle theory files for more information.

\subsection{Casting functions to vectors and matrices}

We now define the functions in Isabelle that will allow us to cast any function on states to the bounded continuous function type, and any function on pairs of states to the square matrix type: 

\begin{lstlisting}[basicstyle=\footnotesize\ttfamily, mathescape = true]{}
definition vectify :: "('s $\Rightarrow$ real) $\Rightarrow$ ('s, real) bcontfun" where
"vectify f = Bcontfun f"

definition vectify_inv :: "('s, real) bcontfun $\Rightarrow$ ('s $\Rightarrow$ real)" 
  where
"vectify_inv v = apply_bcontfun v"

lemma "vectify (vectify_inv v) = v"

lemma "vectify_inv (vectify f) = f"

definition matrify :: "('s $\Rightarrow$ 's $\Rightarrow$ real) $\Rightarrow$ 's sq_mtx" where
"matrify f = to_mtx ($\chi$ i j. f i j)"

definition matrify_inv :: "'s sq_mtx $\Rightarrow$ ('s $\Rightarrow$ 's $\Rightarrow$ real)" where
"matrify_inv M = ($\lambda$i j. vec_nth (sq_mtx_ith M i) j)"

lemma "matrify (matrify_inv A) = A"

lemma "matrify_inv (matrify f) = f"
\end{lstlisting}

We note that \texttt{vectify (Vexpected p)} is the vector of the expected reward under policy \texttt{p} against the states. Using \texttt{vectify}, we define another vector which holds the expected reward for taking a single step from a state under policy $p$, using the function \texttt{single\_reward\_vec}:

\begin{lstlisting}[basicstyle=\footnotesize\ttfamily, mathescape = true]{} 
definition single_reward_vec :: "('s $\Rightarrow$ 's pmf) $\Rightarrow$ 's $\Rightarrow$ real" 
  where
"single_reward_vec p = ($\lambda$s. RQslice 0 p s (p s))"
\end{lstlisting}

\noindent and using \texttt{matrify} we define a matrix which forms the transition matrix under policy \texttt{p}, using a function called \texttt{transition\_matrix}. We then show that this transition matrix is a stochastic matrix:\

\begin{lstlisting}[basicstyle=\footnotesize\ttfamily, mathescape = true]{} 
definition transition_matrix :: "('s $\Rightarrow$ 's pmf) $\Rightarrow$ 's sq_mtx" 
  where
"transition_matrix p = matrify ($\lambda$s s'. pmf (p s) s')"

theorem trans_matr_stochastic:
  fixes p :: "'s $\Rightarrow$ 's pmf"
  assumes "p$\in$policies"
  shows "stochastic_mtx (transition_matrix p)"
\end{lstlisting}

We have now built the concepts needed for our proof of the existence of an optimal policy. The groundwork thus laid, we go on to the proof itself.

\subsection{An optimal policy definition}

Define $P$ as the set of all valid policies. As we have noted before, this is a finite set as we have finite states and finite actions we can take from each one. Define $v_\pi$ as the vector of the function which returns the value (the total expected discounted reward) of each state under policy $\pi$ (see section \ref{sec:Background}). Then define $v^*$ as the supremum in the vector space $V$ of the set $\{v_\pi:\pi\in P\}$.\ 

Recall that the supremum here is the least vector in $V$ such that it is greater than or equal to every $v_\pi$. Note that even though the set of policies is finite, we cannot guarantee that the supremum $v^*$ is attained as $V$ is only partially ordered. We \textit{can} say that:\ 
$$\forall s. \exists \pi. \, v^*(s) = \sup_{\pi\in P}\{v_\pi(s)\} = \max_{\pi\in P}\{v_\pi(s)\}$$ 

So we know that for each component $s$ of $v^*$, there exists at least one $v_\pi$ such that its component $s$ attains the value $v^*(s)$. We just cannot yet say that there is one $v_\pi$ that does so for every $s$. Indeed, \textit{this} is what we are trying to prove. \

If we can find a $\pi^*$ such that $v_{\pi^*} = v^*$, then we have a universally optimal policy for all starting states, as we will know that for any other policy, the $V$ value of any state must be less than or equal to that of $v_{\pi^*}$.\

We begin to formalise this by defining the optimum policy on state-action pairs via a function \texttt{optimal\_policies} that returns these as a set for a given state \texttt{s} and action \texttt{a}. We also define the optimum policy on states using a function \texttt{optimal\_policies\_state} that returns these for a given \texttt{s}. We demonstrate that these both exist for any choice of state or valid state-action pair. These will not form part of the definition of a universal optimal policy directly, but to demonstrate that our definition works we will want to show how it relates to the universal optimal policy conditions. In order to see these relationships, we also refer the reader back to our definitions of \texttt{Qexpected} and \texttt{Vexpected} in section \ref{subsec:expected}.\

\begin{lstlisting}[basicstyle=\footnotesize\ttfamily, mathescape = true]{}
definition optimal_policies :: "'s $\Rightarrow$ 's pmf $\Rightarrow$ ('s $\Rightarrow$ 's pmf) set" 
  where
"optimal_policies s a = 
 {p$\in$policies. 
   ($\forall$p'$\in$policies. Qexpected p' s a $\leq$ Qexpected p s a)}"

definition optimal_policies_state :: "'s $\Rightarrow$ ('s $\Rightarrow$ 's pmf) set" 
  where
"optimal_policies_state s = 
  {p$\in$policies. 
    ($\forall$p'$\in$policies. Qexpected p' s (p' s) $\leq$ Qexpected p s (p s))}"
\end{lstlisting}

Next, we define \texttt{Vmax}, a function that returns the highest possible $V$ value for a state \texttt{s}. We prove that \texttt{p} is in \texttt{optimal\_policies\_state s} if and only if \texttt{Vexpected p s = Vmax s}:\

\begin{lstlisting}[basicstyle=\footnotesize\ttfamily, mathescape = true]{}
definition Vmax :: "'s $\Rightarrow$ real" where
"Vmax s = Max ($\bigcup$p$\in$policies.{Vexpected p s})"

lemma Vmax_optimal:
  fixes s :: 's and p :: "('s $\Rightarrow$ 's pmf)"
  assumes p_in_pol:"p$\in$policies"
  shows "(p$\in$optimal_policies_state s) = (Vexpected p s = Vmax s)"
\end{lstlisting}

We next define the set of universally optimal policies as the set of policies whose expected reward vector is greater than or equal to that of all other policies (recalling the ordering that we have on these vectors from section \ref{subsec:bcfs}). We show that this definition is equivalent to the intersection of the sets of optimal policies over the individual states, as we would expect, and equivalent to defining it as the set of policies where \texttt{Vexpected p = Vmax}.

\begin{lstlisting}[basicstyle=\footnotesize\ttfamily, mathescape = true]{}
definition optimal_policies_universal :: "('s $\Rightarrow$ 's pmf) set" 
  where
"optimal_policies_universal =
 {p$\in$policies. 
  ($\forall$p'$\in$policies. 
    vectify (Vexpected p) $\geq$ vectify (Vexpected p'))}"

theorem universal_equivalence:
  shows "optimal_policies_universal =
             ($\bigcap$(range optimal_policies_state))"
    
lemma universal_optimal_policy_Vmax_equivalence:
  fixes p :: "'s $\Rightarrow$ 's pmf"
  assumes "p$\in$policies"
  shows "(Vexpected p = Vmax) = (p$\in$optimal_policies_universal)"
\end{lstlisting}

\subsection{The $L_\pi$ operator and its supremum, $\mathcal{L}_{max}$}
\label{subsec:Loperator}

For each policy $\pi$, define $r_\pi(s)$ as the function which returns the expected immediate reward an agent earns for taking the action $\pi(s)$ from $s$. Recall that $T^\pi$ is the transition matrix for any $\pi$. Using this, we derive a vector form of Bellman's equation (see section \ref{eqn:bellman}):\

\begin{align}
\label{eqn:vectorbell}
    v_\pi = r_\pi + \gamma T^\pi v_\pi
\end{align}

We prove this in Isabelle by assuming that we have convergence of value, which we have previously proven when \texttt{$\gamma<1$} or when we have inevitable terminating states (see section \ref{subsec:expected}):

\begin{lstlisting}[basicstyle=\footnotesize\ttfamily, mathescape = true]{} 
theorem vec_expr_value:
  fixes p :: "'s $\Rightarrow$ 's pmf"
  assumes "p$\in$policies" 
    and "$\bigwedge$s. $\forall$a$\in$K s. convergent ($\lambda$n. Qexpectedn n p s a)"
  shows "vectify (Vexpected p) = vectify (single_reward_vec p) 
    + $\gamma$ $*_R$ (transition_matrix p) $*_M$ (vectify (Vexpected p))"
\end{lstlisting}
\noindent Note the use of \texttt{vectify} here, needed to so that we can have our expected values in vector form. 

Next, we suppose that $u,v\in V$ and $\gamma<1$ and show the following, which will be needed for subsequent proofs:\

\begin{align}
    u\geq 0 & \implies \forall\pi. (I - \gamma T^\pi)^{-1}u \geq u\\
    \label{eqn:uvrel}
    u\geq v & \implies \forall\pi. (I - \gamma T^\pi)^{-1}u \geq (I - \gamma T^\pi)^{-1}v
\end{align}

\noindent In Isabelle, they are formalised as follows:\

\begin{lstlisting}[basicstyle=\footnotesize\ttfamily, mathescape = true]{} 
lemma discounted_transition_inverse_incr_gezero:
  fixes p :: "'s $\Rightarrow$ 's pmf" and u :: "('s, real) bcontfun"
  assumes "p$\in$policies" and "$\gamma$<1" and "u $\geq$ 0"
  shows "((1 - $\gamma$ *$_R$ (transition_matrix p))$^{-1}$ *$_M$ u) $\geq$ u"
  
lemma discounted_transition_inverse_incr_gevec:
  fixes p :: "'s $\Rightarrow$ 's pmf" and u v :: "('s, real) bcontfun"
  assumes "p$\in$policies" and "$\gamma$ < 1" and "u $\geq$ v"
  shows "((1 - $\gamma$ *$_R$ (transition_matrix p))$^{-1}$ *$_M$ u) $\geq$ 
    ((1 - $\gamma$ *$_R$ (transition_matrix p))$^{-1}$ *$_M$ v)"
\end{lstlisting}

\noindent The proofs here depend on the fact that our the vector space is a Banach space and on the invertability properties of matrices with a spectral radius less than 1, which we proved previously (see section \ref{subsec:gelfform}).\

Next, define the operator $L_\pi$ on any vector $v$ as $L_\pi v = r_\pi + \gamma T_\pi v$. Using  (\ref{eqn:vectorbell}), it is clear that $v_\pi$ is a fixed point (but not necessarily a unique fixed point -- we prove this below) for $L_\pi$ as $L_\pi v_\pi = v_\pi$. In Isabelle, we formalise the \texttt{L} function as:\

\begin{lstlisting}[basicstyle=\footnotesize\ttfamily, mathescape = true]{} 
definition L :: "('s $\Rightarrow$ 's pmf) $\Rightarrow$ ('s, real) bcontfun 
  $\Rightarrow$ ('s, real) bcontfun" where
"L p v = vectify (single_reward_vec p) +
             $\gamma$ $*_R$ ((transition_matrix p) $*_M$ v)"
\end{lstlisting}

We then show that, using Gelfand's formula and (\ref{eqn:specradinv}), that $v_\pi$ is the unique fixed point for $L_\pi$. We go on to derive the following equation assuming $\gamma<1$ and using (\ref{eqn:uvrel}) and the fact that the matrix space is complete:\

\begin{align}
\label{eqn:inversevpi}
    v_\pi = (I - \gamma T^\pi)^{-1}r_\pi
\end{align}

This shows that provided we can find the inverse of $(I - \gamma T^\pi)$, which (\ref{eqn:specradinv}) guarantees when we have $\gamma<1$ and thus a spectral radius less than one, we can calculate the vector of $V$ values for a given policy using only its transition matrix and expected rewards.\

We formalise these results in Isabelle:\

\begin{lstlisting}[basicstyle=\footnotesize\ttfamily, mathescape = true]{} 
theorem find_expected_value:
  fixes p :: "'s $\Rightarrow$ 's pmf"
  assumes "p$\in$policies" and "$\gamma$<1" 
  shows "vectify (Vexpected p) = 
    (1 - ($\gamma$ $*_R$ (transition_matrix p)))$^{-1}$ 
      $*_M$ (vectify (single_reward_vec p))"
\end{lstlisting}

Now, define the operator $\mathcal{L}_{max}$ over vectors in $V$ as follows (note that in Puterman's text, this is denoted $\mathcal{L}$ -- we add the subscript here for clarity):\

\begin{align}
    \mathcal{L}_{max}v = \sup_\pi\{L_\pi v\}
\end{align}

There is some confusing wording in Puterman beneath this definition, where there is an attempt to clarify the definition but it is discernible from the context that this supremum is intended to be over the vector space $V$ with respect to its partial ordering. Recall once more that $V$ is partially ordered, so we cannot be sure that $\mathcal{L}_{max}v$ is attained by any one choice of policy $\pi$. We know, again given our finite choice of states, that for any single component $s$ of $\mathcal{L}_{max}v$, there is a choice of $\pi$ such that $(L_\pi v)(s) = (\mathcal{L}_{max}v)(s)$, but we do not yet know that there is always a policy $\pi$ such that $L_\pi v = \mathcal{L}_{max}v$.\

We now proceed to define an \texttt{$\mathcal{L}_{max}$} function in Isabelle by taking the maximum values for \texttt{L} across the policies, and we prove that for any \texttt{v} it is attained by some policy \texttt{p}:\

\begin{lstlisting}[basicstyle=\footnotesize\ttfamily, mathescape = true]{} 
definition $\mathcal{L}_{max}$ :: 
  "('s, real) bcontfun $\Rightarrow$ ('s, real) bcontfun" where
"$\mathcal{L}_{max}$ b = Bcontfun ($\lambda$s. 
  Max($\bigcup$p$\in$policies. {apply_bcontfun (L p b) s}))"

theorem $\mathcal{L}_{max}$_attained:
  shows "$\exists$p$\in$policies. L p v = $\mathcal{L}_{max}$ v"
\end{lstlisting}

Note that Puterman does not prove that $\mathcal{L}_{max}$ is attained immediately in his proof -- rather it is his final step, under a set of assumptions we do not need to consider (as we deal with the finite MDP case only). We prove it now to fix an issue with one of Puterman's proofs as we will detail soon.\

Next is the biggest step toward the proof of the existence of an optimal policy. We show using  (\ref{eqn:uvrel}) that (assuming $0\leq\gamma<1$):\

\begin{align}
\label{eqn:veqv*}
\mathcal{L}_{max}v = v \implies v = v^*
\end{align}

There is actually a minor error in Puterman's proof here that we uncovered during our formalisation -- this is the reason why we show that $\mathcal{L}_{max}$ is attained earlier than Puterman does. During a particular part of his proof, Puterman asserts:\

\begin{align}
v \leq \mathcal{L}_{max}v \implies \forall \epsilon > 0 . \exists \pi . \forall s . v(s) \leq L_\pi v(s) + \epsilon
\end{align}

\noindent He uses the $\epsilon$ here as he is also considering the infinite state case (where the supremum may not be attained even on a componentwise basis); it can be dismissed in the finite case, leaving:

\begin{align}
\label{eqn:wrongproof}
v \leq \mathcal{L}_{max}v \implies \exists \pi . v(s) \leq L_\pi v(s)
\end{align}

As we have previously discussed, it has not yet been shown that there exists a single policy for which this can be true. Puterman goes on to prove this much later as the final step in his optimal policy existence proof (albeit only under certain assumptions -- one choice of which is that the MDP is finite). Thankfully his delayed proof does not rely on any of his theorems where he seems to have implicitly assumed it. In any case, Puterman does not mention the limiting assumptions in his proof of (\ref{eqn:wrongproof}) here, and this proof is therefore incomplete.\

We formally prove in Isabelle that if there is a fixed point for \texttt{$\mathcal{L}_{max}$}, that it must be \texttt{vectify (Vmax)}:\

\begin{lstlisting}[basicstyle=\footnotesize\ttfamily, mathescape = true]{} 
theorem v_eq_$\mathcal{L}_{max}$_v:
  fixes v :: "('s, real) bcontfun"
  assumes "v = $\mathcal{L}_{max}$ v" "$\gamma$<1"
  shows "v = vectify (Vmax)"
\end{lstlisting}

\subsection{Final steps}
\label{subsec:final}

We then prove that (under the assumption $\gamma<1$) both $L_\pi$ and $\mathcal{L}_{max}$ are contraction mappings, in other words for any $u,v\in V$, $\lVert L_\pi u - L_\pi v \rVert \leq \lVert u - v\rVert$ and likewise for $\mathcal{L}_{max}$. This allows us to use the Banach fixed point theory (along with our proof that V is a Banach space) to show that there exists a unique fixed point of both operators. In Isabelle, we have:\

\begin{lstlisting}[basicstyle=\footnotesize\ttfamily, mathescape = true]{} 
theorem contraction_L:
  fixes u v :: "('s, real) bcontfun" and p :: "'s $\Rightarrow$ 's pmf"
  assumes "$\gamma$<1" "p$\in$policies"
  shows "dist (L p u) (L p v) $\leq$ $\gamma$ * dist u v"

theorem contraction_$\mathcal{L}_{max}$:
  fixes u v :: "('s, real) bcontfun"
  assumes "$\gamma$<1"
  shows "dist ($\mathcal{L}_{max}$ u) ($\mathcal{L}_{max}$ v) $\leq$ $\gamma$ * dist u v"
  
theorem unique_solution_L:
  fixes v :: "('s, real) bcontfun" and p :: "'s $\Rightarrow$ 's pmf"
  assumes "$\gamma$<1" "p$\in$policies"
  shows "$\exists$!v. L p v = v"

theorem unique_solution_$\mathcal{L}_{max}$:
  fixes v :: "('s, real) bcontfun" and p :: "'s $\Rightarrow$ 's pmf"
  assumes "$\gamma$<1" "p$\in$policies"
  shows "$\exists$!v. $\mathcal{L}_{max}$ v = v"
\end{lstlisting}

All the pieces of the proof are in place now and the only remaining significant step is to show that there exists a policy $\pi$ such that for a choice of $v$, $L_\pi v = \mathcal{L}_{max} v$. We do this by noting that for each choice of state component, $s$ say, we can maximise the value for $L_\pi v(s)$ just by assuming that $\pi$ takes a particular choice of action from that state; we then note that if we make that assumption for $\pi$ across every state, that $L_\pi v=\mathcal{L}_{max}v$. Note that we have already shown this in Isabelle with the \texttt{$\mathcal{L}_{max}$\_attained} theorem in the previous section.\

This done, we have found a policy $\pi^*$ such that $L_{\pi^*} v = \mathcal{L}_{max} v$. $\mathcal{L}_{max}$ has a unique fixed point $v^*$ (\ref{eqn:veqv*}), which by our vector version of Bellman's equation (\ref{eqn:vectorbell}), we know must also be the unique fixed point for $L_{\pi^*}$, and hence the expected value function for policy $\pi^*$.\

By our definition of $v^*$, we thus know that $\pi^*$ attains the highest possible expected value for each state component, and is the universally optimal policy.\

We demonstrate this in Isabelle by showing that we can select a policy \texttt{p} such that it attains \texttt{$\mathcal{L}_{max}$} on the vector of its expected values on the states using the \texttt{$\mathcal{L}_{max}$\_attained} theorem. We then know that \texttt{$\mathcal{L}_{max}$ (vectify (Vexpected p)) = (vectify (Vexpected p))} and therefore that \texttt{vectify (Vexpected p) = Vmax}. This all leads to a proof of the very satisfying statement of our sought-after theorem:

\begin{lstlisting}[basicstyle=\footnotesize\ttfamily, mathescape = true]{} 
theorem universal_optimal_policy_exists:
  assumes "$\gamma$<1"
  shows "$\exists$p. p$\in$optimal_policies_universal"
\end{lstlisting}

\section{Value and Policy Iteration}
\label{sec:iterations}

As the next step in our formalisation, we introduce the value iteration and policy iteration algorithms, and formally prove that they work as intended. Again, our work here focuses on formalising the work presented in Puterman's textbook \cite{puterman2014markov}.\

\subsection{Finding an optimal policy}

In the previous section we have proved that a universal optimal policy always exists on any finite MDP with a discount. We know this is important because an agent following this policy should behave optimally over the MDP, attaining maximal reward.\

But how do we construct the optimal policy for an arbitrary MDP? In the situation where we do not know the structure of the MDP and must discover it by exploration, reinforcement learning is the typical process used. However, we have simpler methods available in the case where we know the properties of the MDP, its states, actions, rewards and the transition probabilities associated with it. These methods include value iteration and policy iteration, two algorithms intended to compute the optimal policy, or an arbitrarily close approximation in the case of value iteration, in finite time.\

In this section we will discuss each algorithm, before going on to show their representation in Isabelle/HOL, and our proofs of their correctness.\

\subsection{Value iteration}
\label{subsec:valueiter}

The first algorithm we will look at is value iteration. Before we examine it, we will define an $\epsilon$-optimal policy as one where the distance between its expected value vector and that of an optimal policy is at most $\epsilon$. This distance is defined using the vector norm, so we are saying that a policy $p_\epsilon$ is $\epsilon$-optimal if and only if:\ 

$$\forall s. \lvert V_{p_\epsilon}(s) - V_*(s) \rvert \leq \epsilon$$

The value iteration algorithm constructs an $\epsilon$-optimal policy in finite time, via the following process:\

\begin{enumerate}
    \item Choose an arbitrary vector $v_0$ in $V$, the vector space of bounded functions on the states. Choose an $\epsilon$ arbitrarily close to 0. We begin with time step $n=0$.
    \item Find $v_{n+1}$ using:
    $$v_{n+1} = \mathcal{L}_{max} v_n$$
    \item If
    $$ \lVert v_{n+1}-v_n \rVert \leq \epsilon (1-\gamma) / 2 \gamma$$
    then proceed, otherwise increment $n$ and repeat step 2.
    \item For each $s$, find $p_\epsilon(s)$ using:
    $$p_\epsilon(s) = \argmax_{a\in A(s)} \bigg\{ \sum_{s'\in a} P(s, a, s') \big( R(s,s',a) + \gamma v_{n+1}(s) \big) \bigg\}$$
\end{enumerate}

\noindent The policy $p_\epsilon$ found this way is $\epsilon$-optimal.\medskip

We represent this algorithm in Isabelle via four functions. The first one, \texttt{Value\_Iteration}, gives us the vector $v_n$ found in step 2 of the algorithm. It takes as parameters the time step of the algorithm \texttt{n} and the chosen initial vector \texttt{v}.\

\begin{lstlisting}[basicstyle=\footnotesize\ttfamily, mathescape = true]{} 
fun Value_Iteration :: 
  "nat $\Rightarrow$ ('s, real) bcontfun $\Rightarrow$ ('s, real) bcontfun" where
 "Value_Iteration 0 v = v" 
|"Value_Iteration (Suc n) v = $\mathcal{L}_{max}$ (Value_Iteration n v)"
\end{lstlisting}

\texttt{Value\_Iteration\_Algo\_n} gives us the first time step $n$ when the conditions for step 3 of the algorithm are met, using Isabelle's \texttt{LEAST} operator. To find this, we must specify the parameter \texttt{$\epsilon$} instead of a time step.\

\begin{lstlisting}[basicstyle=\footnotesize\ttfamily, mathescape = true]{} 
definition Value_Iteration_Algo_n :: 
  "real $\Rightarrow$ ('s, real) bcontfun $\Rightarrow$ nat" where
"Value_Iteration_Algo_n $\epsilon$ v = 
  (LEAST n. dist (Value_Iteration (Suc n) v) (Value_Iteration n v) 
    < ($\epsilon$ * (1 - $\gamma$))/ (2 * $\gamma$))"
\end{lstlisting}

\texttt{Value\_Iteration\_Algo\_v} gives us the vector associated with the time step we need to calculate our policy.\

\begin{lstlisting}[basicstyle=\footnotesize\ttfamily, mathescape = true]{} 
definition Value_Iteration_Algo_v :: 
  "real $\Rightarrow$ ('s, real) bcontfun $\Rightarrow$ ('s, real) bcontfun" where
"Value_Iteration_Algo_v $\epsilon$ v = 
    Value_Iteration ((Value_Iteration_Algo_n $\epsilon$ v)+1) v"
\end{lstlisting}

Finally, \texttt{Value\_Iteration\_Algo} constructs the policy associated with completion of the algorithm.\

\begin{lstlisting}[basicstyle=\footnotesize\ttfamily, mathescape = true]{} 
definition Value_Iteration_Algo :: 
  "real $\Rightarrow$ ('s, real) bcontfun $\Rightarrow$ ('s $\Rightarrow$ 's pmf)" where
"Value_Iteration_Algo $\epsilon$ v = 
 ($\lambda$s. arg_max 
   ($\lambda$a. ($\sum$s'$\in$a. pmf a s' * (R(s, s', a) +  
     $\gamma$ * apply_bcontfun (Value_Iteration_Algo_v $\epsilon$ v) s'))) 
   ($\lambda$a. a$\in$K s))"
\end{lstlisting}

To show that our representations are accurate, we prove basic properties of the value iteration algorithm and conclude by formally verifying that it leads to an $\epsilon$-optimal policy. We begin by proving that the vector produced by iteration tends towards the optimal value vector as $n\to\infty$. This proof falls out of the use of the \texttt{$\mathcal{L}_{max}$} operator in our definitions and our previous proof of the existence of an optimal policy.\

\begin{lstlisting}[basicstyle=\footnotesize\ttfamily, mathescape = true]{} 
lemma Value_Iteration_Converges:
  fixes v :: "('s, real) bcontfun"
  assumes "$\gamma$<1"
  shows "lim ($\lambda$n. Value_Iteration n v) = vectify Vmax"
\end{lstlisting}

Next we show that provided $\epsilon>0$, we can always find a finite $n$ that fulfils the conditions of step 3 of the value iteration algorithm. This falls easily out of the limit property we have just proven and the fact that our vectors form a Banach space. Note that we must specify an additional assumption, that $\gamma>0$ here, as otherwise division by zero would be possible on the value we test against.

\begin{lstlisting}[basicstyle=\footnotesize\ttfamily, mathescape = true]{} 
lemma Value_Iteration_Algo_n_Exists:
  fixes v :: "('s, real) bcontfun" and $\epsilon$ :: real
  assumes "$0<\gamma$" "$\gamma$<1" 
  shows "$\forall\epsilon$>0. $\exists$n. 
          dist (Value_Iteration (Suc n) v) (Value_Iteration n v) <
            ($\epsilon$ * (1 - $\gamma$))/ (2 * $\gamma$)"
\end{lstlisting}

We continue by proving that the final result of the algorithm is always a policy. This is important in our formalisation as policies are not a type -- they are members of a set with certain properties. So if we have a function that computes a result of the correct type, we need to be sure that it meets the conditions for being in the set to know that it is well-behaved. This proof is not completely trivial -- we show that the \texttt{arg\_max} used in the definition always exists in the set of valid actions on a state, and then show that the resulting function always produces such a valid action.\

\begin{lstlisting}[basicstyle=\footnotesize\ttfamily, mathescape = true]{} 
lemma Value_Iteration_Algo_policy:
  fixes $\epsilon$ :: real and v :: "('s, real) bcontfun"
  shows "Value_Iteration_Algo $\epsilon$ v $\in$ policies"
\end{lstlisting}

This next proof is crucial to establishing $\epsilon$-optimality: we show that the policy $p_\epsilon$ we produce with \texttt{Value\_Iteration\_Algo} and the vector $v_\epsilon$ we produce with \texttt{Value\_Iteration\_Algo\_v} gives the equivalence $L_{p_\epsilon} (v_\epsilon) = \mathcal{L}_{max} (v_\epsilon)$.

\begin{lstlisting}[basicstyle=\footnotesize\ttfamily, mathescape = true]{} 

lemma Value_Iteration_Algo_Equiv:
fixes $\epsilon$ :: real and v :: "('s, real) bcontfun"
shows "L (Value_Iteration_Algo $\epsilon$ v) (Value_Iteration_Algo_v $\epsilon$ v) = 
         $\mathcal{L}_{max}$ (Value_Iteration_Algo_v $\epsilon$ v)"
\end{lstlisting}

Our final proof here makes use of the equivalence we have established and the triangle inequality for norms to give an upper bound on the distance between the value vector produced by our constructed policy and the optimal value vector. The equivalence is necessary to be able to use the contraction mapping properties of the $\mathcal{L}_{max}$ and $L_{p_\epsilon}$ operators to establish this bound. Once we have the bound, simple algebraic manipulation and repeated use of the triangle inequality leads to the result.\

\begin{lstlisting}[basicstyle=\footnotesize\ttfamily, mathescape = true]{} 
theorem Value_Iteration_Epsilon_Optimal:
  fixes $\epsilon$ :: real and v :: "('s, real) bcontfun"
  assumes "$\gamma$>0" "$\gamma$<1" "$\epsilon$>0"
  shows "dist (vectify (Vexpected (Value_Iteration_Algo $\epsilon$ v))) 
            (vectify Vmax) < $\epsilon$"
\end{lstlisting}

We conclude, then, by demonstrating that the policy we have produced has a value vector less than $\epsilon$ distant from the optimal value vector -- in other words, that it is $\epsilon$-optimal as expected.\

\subsection{Policy iteration}

We move on to examine policy iteration. Unlike value iteration, policy iteration constructs a universally optimal policy in finite time, not simply an $\epsilon$-optimal policy. We will formally prove that this is achieved by the algorithm. 

Policy iteration follows this process:\

\begin{enumerate}
    \item Choose an arbitrary policy $p_0$. We begin with time step $n=0$.
    \item Find $v_n$ using:
    $$v_n = (I-\gamma T_{p_n})^{-1} r_{p_n}$$
    Recall that $r_{p_n}$ is the vector of expected rewards earned by taking a single step from a state and $T_{p_n}$ is the transition matrix associated with policy $p_n$.
    \item Select a policy $p_{n+1}$ such that:
    $$p_{n+1}(s) = \argmax_{a\in A(s)} \bigg\{ \sum_{s'\in a} P(s, a, s') \big( R(s,s',a) + \gamma v_n(s) \big) \bigg\}$$
    If $p_n$ can fulfil this condition, choose it.
    \item If $p_{n+1}=p_n$ then call it $p^*$ and conclude the process. Otherwise, increment $n$, return to step 2 and continue.
\end{enumerate}

The policy $p^*$ found this way is universally optimal.\medskip

As we did for value iteration, we represent this algorithm in Isabelle with a number of functions.\

\texttt{Policy\_Iteration} and \texttt{Policy\_Iteration\_v} are mutually recursive functions that give us the result of step 3 of the algorithm. \texttt{Policy\_Iteration\_v} represents the vector $v_n$ found in step 2. They take as parameters the time step \texttt{n} of the algorithm and the chosen initial policy \texttt{p}. Note that we use an \texttt{if} conditional to check if the policy constructed in the previous increment meets the conditions for the maximisation; if it does we select it, otherwise we choose an arbitrary policy that meets the conditions.\

\begin{lstlisting}[basicstyle=\footnotesize\ttfamily, mathescape = true]{} 
fun Policy_Iteration :: "nat $\Rightarrow$ ('s $\Rightarrow$ 's pmf) $\Rightarrow$ ('s $\Rightarrow$ 's pmf)"
  and Policy_Iteration_v :: 
    "nat $\Rightarrow$ ('s $\Rightarrow$ 's pmf) $\Rightarrow$ ('s, real) bcontfun" where
  "Policy_Iteration 0 p = p" 
| "Policy_Iteration_v n p 
  = (1 - $\gamma$ *$_R$ (transition_matrix (Policy_Iteration n p)))$^{-1}$ *$_M$
     (vectify (single_reward_vec (Policy_Iteration n p)))" 
| "Policy_Iteration (Suc n) p = 
    (if ($\forall$s. is_arg_max ($\lambda$a. 
      ($\sum$s'$\in$a. pmf a s' * (R(s, s', a) +
         $\gamma$ * apply_bcontfun (Policy_Iteration_v n p)  s'))) 
      ($\lambda$a. a$\in$K s) ((Policy_Iteration n p) s))
     then (Policy_Iteration n p) 
     else ($\lambda$s. arg_max ($\lambda$a. 
            ($\sum$s'$\in$a. pmf a s' * (R(s, s', a) +
               $\gamma$ * apply_bcontfun (Policy_Iteration_v n p) s'))) 
            ($\lambda$a. a$\in$K s)))"
\end{lstlisting}

\texttt{Policy\_Iteration\_Algo\_n} gives us the first time step at which the conditions for step 4 are met, similarly to the corresponding value iteration function, and likewise using Isabelle's \texttt{LEAST} operator.\

\begin{lstlisting}[basicstyle=\footnotesize\ttfamily, mathescape = true]{} 
definition Policy_Iteration_Algo_n :: "('s $\Rightarrow$ 's pmf) $\Rightarrow$ nat" 
  where
"Policy_Iteration_Algo_n p =
  (LEAST n. (Policy_Iteration (Suc n) p = Policy_Iteration n p))"
\end{lstlisting}

\noindent and finally, \texttt{Policy\_Iteration\_Algo} gives us the policy $p^*$ which is the final outcome of the algorithm.\

\begin{lstlisting}[basicstyle=\footnotesize\ttfamily, mathescape = true]{} 
definition Policy_Iteration_Algo :: "('s $\Rightarrow$ 's pmf) 
  $\Rightarrow$ ('s $\Rightarrow$ 's pmf)" where
"Policy_Iteration_Algo p = 
  Policy_Iteration (Policy_Iteration_Algo_n p) p"
\end{lstlisting}

As before, we need to verify some basic properties of the algorithm to show our representation is correct before moving on to our proof of optimality.\

We begin by showing that for every step of the algorithm produces a new policy $p_n$, which meets the set-based definition of policy we are using in our model. Again, we simply have to show that the \texttt{arg\_max} function can find a valid action and then show that these choices of actions define a policy.\

\begin{lstlisting}[basicstyle=\footnotesize\ttfamily, mathescape = true]{} 
lemma Policy_Iteration_policy:
  fixes p :: "'s $\Rightarrow$ 's pmf" and n :: nat
  assumes "p$\in$policies"
  shows "Policy_Iteration n p $\in$ policies"
\end{lstlisting}

Next we show that for subsequent policies there is an equivalence such that $L_{p_n+1} v_{p_n}=\mathcal{L}_{max} v_{p_n+1}$. Again, this is crucial for subsequent proofs, but unlike value iteration, we do not use the contraction mapping properties of these functions, but rather their fixed points -- we are not showing an approximation to optimality, but optimality itself.\

\noindent \begin{minipage}{\linewidth}
\begin{lstlisting}[basicstyle=\footnotesize\ttfamily, mathescape = true]{} 
lemma Policy_Iteration_Equiv:
  fixes p :: "'s $\Rightarrow$ 's pmf" and n :: nat
  assumes "p$\in$policies" "$\gamma$<1"
  shows "L (Policy_Iteration (Suc n) p) 
           (vectify (Vexpected (Policy_Iteration n p))) =
                $\mathcal{L}_{max}$ (vectify (Vexpected (Policy_Iteration n p)))"
\end{lstlisting}
\end{minipage}

A crucial part of our optimality proof is to show that every time step of the algorithm (unless there is no change) improves the policy -- every policy constructed has a value vector that is strictly greater than that of the previous step. This proof relies on manipulation of vector calculations using the invertability properties of the matrices that are used in the policy iteration definitions.\

\begin{lstlisting}[basicstyle=\footnotesize\ttfamily, mathescape = true]{} 
lemma Policy_Iteration_inc:
  fixes p :: "'s $\Rightarrow$ 's pmf" and n :: nat
  assumes "p$\in$policies" "$\gamma$<1"
  shows "vectify (Vexpected (Policy_Iteration (Suc n) p)) $\geq$
            vectify (Vexpected (Policy_Iteration n p))"
\end{lstlisting}

Our next proof demonstrates that if we have a step of the algorithm such that the policy does not change, that the expected value vector of that policy must be optimal -- the policy is universally optimal. This makes use of the \texttt{Policy\_Iteration\_Equiv} lemma and the fixed points of the $L_{p_n}$ and $\mathcal{L}_{max}$ operators.\

\begin{lstlisting}[basicstyle=\footnotesize\ttfamily, mathescape = true]{} 
lemma Policy_Iteration_Works:
  fixes p :: "'s $\Rightarrow$ 's pmf" and n :: nat
  assumes "p$\in$policies" "$\gamma$<1" 
          "Policy_Iteration (Suc n) p = Policy_Iteration n p"
  shows 
  "vectify (Vexpected (Policy_Iteration n p)) = vectify (Vmax)"
\end{lstlisting}

We know that if the policy does not remain the same when the algorithm iterates, that it improves. And we know that if the policy does remains the same, then it must be universally optimal. We know that there are only finite many valid policies. It is obvious from these results that in finite time we will produce a universally optimal policy. Unfortunately, despite the intuitive obviousness here, it remains to be shown formally that this will occur in finite time. We do this via some simple results about monotonicity and by incorporating our proof of finite policies.\

\begin{lstlisting}[basicstyle=\footnotesize\ttfamily, mathescape = true]{} 
lemma Policy_Iteration_Finite:
  fixes p :: "'s $\Rightarrow$ 's pmf"
  assumes "p$\in$policies" "$\gamma$<1"
  shows "$\exists$N. (Policy_Iteration N p) $\in$ optimal_policies_universal"
\end{lstlisting}

This proof complete, we have shown that policy iteration constructs a universally optimal policy in finite time.\

\section{Related work}

The most closely related work to our own is that of H\"olzl \cite{holzl2017markov}, which we have discussed at length in section \ref{subs:holzl} as we use it as the starting point for the current mechanisation.\

Another work which covers similar ground to our own is the CertRL work by Vajjha et al.\ \cite{vajjha2020certrl}. This proves the existence of an optimal policy using the same Giry monad approach taken by H\"olzl and some properties related to value and policy iteration. It uses the Coq theorem prover rather than Isabelle.\ 

It only covers some of the work we have performed on value and policy iteration. It proves value iteration in infinite time will find the optimal policy (the first section of our own set of value iteration proofs) and that policy iteration improves the policy with each iteration. We extend these results by showing that policy iteration completes in finite time with an optimal policy. Note that in the Vajjha et al.'s paper there is an assertion that this extended result was proven but we were unable to locate this proof in their Coq theories. We also prove that value iteration achieves $\epsilon$-optimality in finite time, something which CertRL does not attempt.\

The CertRL work proves optimality for value iteration in finite steps, but only against a finite-horizon MDP (in other words, a Markov Decision Process where the agent is limited to a known number of steps). Our own work assumes an infinite horizon for all proofs.\

The question of whether to use the Giry monad (as CertRL and H\"olzl do) or matrices for transition probability calculations touches upon a more fundamental point. The intention of the current work (and the CertRL work) is to provide researchers with a model they can use while working with processes across MDPs (most probably reinforcement learning).\ 

The mathematical background for working with the Giry monad (category theory and measure theory over probability spaces) is substantially less common than the linear algebra background needed to understand and use matrices. Our argument is that by basing our work upon linear algebra, our formalisation is more accessible to a wider section of the research community, especially those interested in reinforcement learning, as well as making some proofs easier to produce (admittedly while requiring more initial work from us).\

\section{Conclusion}
We have built a formal model of a finite MDP in Isabelle and shown that it allows the mechanisation of the the properties one might expect. In particular:\ 

\noindent \begin{minipage}{\linewidth}
\begin{itemize}
    \item We derived Bellman's equation, which describes the expected total value of being in a given state in an MDP, choosing a particular action, and then subsequently pursuing a given policy.
    \item We formalised the circumstances under which, given infinite time, our valuation of rewards earned by an agent converges.
    \item We developed the vector form of Bellman's equation and proved that its results match our scalar version.
    \item We mechanised a vector-based calculation that, providing $\gamma<1$, gives the expected reward for any policy on any state without requiring calculation of an infinite series.
    \item We showed that, providing $\gamma<1$, at least one optimal policy exists that, for any choice of start state or initial action, produces the maximal expected reward if it is pursued.
    \item We demonstrated that the value iteration algorithm produces an $\epsilon$-optimal policy in finite time.
    \item We showed that the policy iteration algorithm produces a universally optimal policy in finite time.
    \item We found a small gap in Puterman's proof of the existence of an optimal policy and explicitly detailed how it is eliminated in the finite case.
\end{itemize}
\end{minipage}\
\medskip

With regards to the mechanisation effort, the work is split across six Isabelle theories. Two of these are heavily reworked versions of existing theories, namely those associated with the square matrices from the ``Matrices for ODEs" entry in the AFP. The remaining four theories establish a relationship between the \texttt{vec} and the \texttt{bcontfun} types and extend the matrix type with stochastic matrices and some spectral radius results. For good measure, they also cover a sample MDP with proofs of some of its properties using our particular formalisation (which we do not describe in the current paper). Altogether, there are almost 7,500 lines of proof script.\

We believe that the work provides a framework that is general enough to use as a basis for formally verifying a number of different systems that use MDPs: our primary interest is in reinforcement learning, but because we have used matrices to represent operators on functions over our MDPs, the vast majority of the existing mathematical literature on MDPs should be amenable to formalisation using our work as its foundations.\

Isabelle was an effective tool for this purpose. The use of the Isar language breaks our proof into human-readable steps, and means that our mechanical proofs can be human-checked. This gives them greater persuasive weight than a more obtuse theorem proving system might have done.\

Difficulties arose when working with types that were present in existing theories but for which we needed different norms or other properties that were fixed by those theories when proving the type could be interpreted as an instance of (for example) a normed vector space. We needed to redefine the type and then introduce our properties.\ 

It seems mathematically natural, for example, to assume that vectors might be measurable using the operator norm or might be measurable using the Euclidean norm; however, in Isabelle, one would need to define two types of vectors with each norm separately, then build functions to translate between each type. The alternative would be to define a norm as an entirely different function and then fail to inherit the large number of related lemmas that already exists.\

If it were possible to show that a type can be interpreted as a normed vector space using multiple norms, this would make it easier to work with the same type for different purposes. Unfortunately, because this isn't currently possible, this made our work more difficult.

However, this is merely a matter of convenience, as our work has demonstrated that by type redefinition it is possible to establish different norms and inherit the proofs against an existing type with a modicum of effort.\

The decision to use matrices to represent probability transitions rather than using the Giry monad was fruitful. The most noticeable example here was our proof that the expected value associated with a policy can be calculated using a matrix inverse multiplication over its immediate reward vector (section \ref{subsec:Loperator}). In general this choice made it much easier to mechanise the many proofs in the literature that rely on the properties of matrices.

\subsection{Future work}

The most obvious way to extend this work would be to extend the model to non-finite MDPs where rewards are bounded. The results would still follow, although they would require different proof methods, notably a different formalisation of the types used in Isabelle. In particular, our use of the vector and matrix types in Isabelle would need to be redefined as both are only valid with finite dimensionality.\

By deriving Bellman's equation and proving that an optimal policy exists, we lay the groundwork for further formalisation. The goal of optimising an MDP is to find the optimal policy, either using dynamic programming methods where we assume that the rewards are known, or via exploration of the MPD, typically by reinforcement learning.\

Formalising these optimisation methods is the obvious next step. In particular, there is much work currently in reinforcement learning, frequently using experimentation to establish results -- making them open to concerns about reproducibility. We plan to build a formal model able to verify various reinforcement learning results with mathematical proof that could provide researchers with the tools to dispel as much uncertainty as possible fully formally. We intend to begin with a formalisation of the proof that Q learning \cite{watkins1992q} converges to the optimal policy.\

\bibliographystyle{spmpsci}
\bibliography{MDPformalpaper}

\end{document}